\documentclass[twocolumn,aps,superscriptaddress,prd,10pt,showpacs,preprintnumbers,nofootinbib]{revtex4-1}
\usepackage{graphicx,slashed,bbold,subfigure,amsmath,amssymb,enumerate}
\usepackage[utf8]{inputenc} 
\usepackage[T1]{fontenc}
\usepackage[english]{babel}
\usepackage{natbib} 
\usepackage[colorlinks=true,citecolor=blue,linkcolor=blue,urlcolor=blue]{hyperref}
\setcitestyle{square, numbers, comma, sort&compress} 
\usepackage{xcolor}
\begin{document}
  
\title{Magnetic field dependent 't Hooft determinant extended Nambu--Jona-Lasinio model} 
\author{João Moreira}
\email{jmoreira@uc.pt}
\affiliation{CFisUC - Center for Physics of the University of Coimbra, Department of Physics, Faculty of Sciences and Technology, University of Coimbra, 3004-516 Coimbra, Portugal}
\author{Pedro Costa}
\email{pcosta@uc.pt}
\affiliation{CFisUC - Center for Physics of the University of Coimbra, Department of Physics, Faculty of Sciences and Technology, University of Coimbra, 3004-516 Coimbra, Portugal}
\author{Tulio E. Restrepo}\email{tulio.restrepo@posgrad.ufsc.br}
\affiliation{Departamento de F\'{\i}sica, Universidade Federal de Santa
  Catarina, Florian\'{o}polis, SC 88040-900, Brazil}
\affiliation{CFisUC - Center for Physics of the University of Coimbra, Department of Physics, Faculty of Sciences and Technology, University of Coimbra, 3004-516 Coimbra, Portugal}

\begin{abstract}
We study the implications of recent lattice QCD results for the magnetic field dependence of the quarks dynamical masses on the 't Hooft determinant extended Nambu--Jona-Lasinio model in the light and strange quark sectors (\emph{up}, \emph{down} and \emph{strange}). The parameter space is constrained at vanishing magnetic field, using the quarks dynamical masses and the meson spectra, whereas at nonvanishing magnetic field strength the dependence of the dynamical masses of two of the quark flavors is used to fit a magnetic field dependence on the model couplings, both the four-fermion Nambu--Jona-Lasinio interaction and the six-fermion 't Hooft flavor determinant. We found that this procedure reproduces the inverse magnetic catalysis, and the strength of the scalar coupling decreases with the magnetic field, while the strength of the six-fermion 't Hooft flavor determinant increases with the magnetic field.
\end{abstract}

\pacs{11.30.Qc,11.30.Rd,12.39.Fe,14.65.-q,21.65.Qr,75.90.+w
}

\maketitle

\section{Introduction}

The effect of strong magnetic fields in strongly interacting matter plays a very important role in several physical contexts such as in heavy ion collisions \cite{Skokov:2009qp,Voronyuk:2011jd,Kharzeev:2007jp}, in astronomic compact object like magnetars \cite{Duncan:1992hi,Kouveliotou:1998ze}, and in the first phases of the early Universe \cite{Vachaspati:1991nm,Enqvist:1993np}. 
Quantum chromodynamics (QCD) is the theory of the strong interaction between quarks and gluons whose phase diagram have been widely studied by lattice QCD (LQCD) simulations and in the context of effective models also in the presence of magnetic fields (see \cite{Andersen:2014xxa} for a review).

At zero temperature, and in the presence of magnetic fields, LQCD and the vast majority of effective models predict magnetic catalysis, which is the increment of the chiral order parameter, the light quark condensate, as the magnetic field, $B$, increases. This phenomenon is originated by the dominant contribution of the lowest Landau-level \cite{Gusynin:1995nb}.
Nevertheless, at finite temperature and magnetic field, apart from a few exceptions, for example Ref. \cite{Fraga:2012fs}, without the insertion of additional mechanisms, the most effective models fail to predict inverse magnetic catalysis (IMC), contradicting LQCD results, where the pseudocritical temperature for the chiral symmetry restoration decreases as $B$ increases \cite{Bali:2011qj,Bali:2012zg,Endrodi:2015oba}. 
Possible reasons for this discrepancy have been given in Refs. \cite{Fukushima:2012kc,Kojo:2012js,Bruckmann:2013oba,Fraga:2012ev}. 
Particularly, in Ref. \cite{Bruckmann:2013oba}, it was argued that IMC comes from the rearrangement of the Polyakov loop induced by the coupling of magnetic field with the sea quarks. 
This kind of backreaction of the Polyakov loop was implemented in the entangled Polyakov--Nambu--Jona-Lasinio (EPNJL) model \cite{Sakai:2010rp}, where the scalar coupling of the Nambu--Jona-Lasinio (NJL) model \cite{Nambu:1961fr,*Nambu:1961tp} is a function of the Polyakov loop. 
Even though this model by itself also fails to reproduce IMC, in Ref. \cite{Ferreira:2013tba}, it was shown that if the pure-gauge critical temperature $T_0$ is fitted to reproduce LQCD data \cite{Bali:2011qj}, then the EPNJL model reproduces IMC.
Later, it was argued that IMC can be reproduced by mimicking asymptotic freedom, which is absent in many effective models and it is one of the most important characteristics of QCD. 
For instance, in Ref. \cite{Ferreira:2014kpa}, the scalar coupling $G(B)$ of the SU(3) version of the NJL model, was fitted to reproduce the LQCD pseudocritical temperatures for the chiral transitions, $T_c(B)$. 
In Refs. \cite{Farias:2014eca,Farias:2016gmy} the scalar coupling is also made explicitly temperature dependent by fitting it to reproduce the LQCD quark condensate \cite{Bali:2011qj}. 
More recently, in Ref. \cite{Endrodi:2019whh}, the constituent quark masses were calculated as a function of the magnetic field, $B$, using LQCD simulations and then the coupling $G(B)$ was set to reproduce those constituent quark masses. 
It is important to remark that this new way to set $G(B)$ was made within the SU(2) version of the Polyakov--Nambu--Jona-Lasinio model using the proper time formalism.

In the present paper we aim to apply the method proposed in \cite{Endrodi:2019whh} to the more laborious SU(3) version of the 't Hooft extended NJL model (which will henceforth be referred to as the NJL model), where now, due to the 't Hooft six fermions interactions that appear in the model, we have to set two couplings, the scalar coupling, $G(B)$, and the six fermions coupling, $\kappa(B)${\footnote {It is interesting to note that, motivated by phenomenology arguments, a temperature \cite{Kunihiro:1989my} and a density dependence \cite{Costa:2004db,Costa:2005cz} of $\kappa$, in the form of decreasing exponentials, were already proposed in order to achieve an effective restoration of axial U$_A$(1) symmetry.}}.
Also, in this application we consider the more general nondegenerate case $m_u\neq m_d\neq m_s$. We are particularly interested in the behavior of $\kappa$, since in previous applications of the 't Hooft extended NJL model, this coupling was kept independent of the magnetic field.

For the fit of the parameters we follow the procedure: first we fit the six parameters of the model, $\Lambda$, $G$, $\kappa$, $m_u$, $m_d$ and $m_s$ at $T=B=0$; then, at $B\neq 0$ the couplings $G$ and $\kappa$ are set to reproduce $M_d$ and $M_s$ magnetized constituent quark masses respectively, calculated in \cite{Endrodi:2019whh}, while the other parameters are kept $B$ independent.  The other constituent quark mass, $M_u$, is an output in our procedure.

The present paper is organized as follows. In Sec. \ref{sec2} we present the SU(3) 't Hooft extended NJL model. In Sec. \ref{sec3} we fit the parameters of the model to reproduce LQCD results \cite{Endrodi:2019whh} and we present the respective analysis of our results. Finally, in Sec. \ref{sec4} we draw our conclusions and final remarks.

\section{The model}\label{sec2}

\subsection{'t Hooft extended NJL model}
The SU(3) version of the NJL model is given by the Lagrangian density \cite{Klevansky:1992qe},
\begin{align}
 \mathcal{L}_{\text{NJL}}=&\overline{\psi}_f\left[\slashed D^{\mu}-\hat{m}_c\right]\psi_f+\mathcal{L}_{\text{sym}}+\mathcal{L}_{\text{det}}\notag\\
\label{lag}
\end{align}
in which the quark sector includes scalar-pseudoscalar and 't Hooft six fermions interactions that models the axial U$_A$(1) symmetry breaking, with $\mathcal{L}_{\text{sym}}$ and $\mathcal{L}_{\text{det}}$ being \cite{Klevansky:1992qe}
\begin{align}
 \mathcal{L}_{\text{sym}}=&G\left[\left(\overline{\psi}_f\lambda_a \psi_f\right)^2+\left(\overline{\psi}_f i\gamma_5\lambda_a \psi_f\right)^2\right],\notag\\
 \mathcal{L}_{\text{det}}=&\kappa\left\lbrace\text{det}_f\left[\overline{\psi}_f\left(1+\gamma_5\right)\psi_f\right]\right.\notag\\
 &+\left.\text{det}_f\left[\overline{\psi}_f\left(1-\gamma_5\right)\psi_f\right]\right\rbrace,
\end{align}
where $\psi_f$ are the quark fields with $f={u,d,s}$, $m_c=\text{diag}_f(m_u,m_d,m_s)$ is the quark current mass matrix, $\lambda_a$ are the Gell-Mann matrices and $G$ and $\kappa$ are coupling constants. 

In the mean field approximation de effective quarks masses are given by the gap equations
\begin{align}
\label{GapEqs}
\left\{
\begin{array}[c]{c}%
M_u=m_u-G \langle \overline{\psi}_u\psi_u\rangle-\kappa \langle \overline{\psi}_d\psi_d\rangle\langle \overline{\psi}_s\psi_s\rangle \\
M_d=m_d-G \langle \overline{\psi}_d\psi_d\rangle-\kappa \langle \overline{\psi}_u\psi_u\rangle\langle \overline{\psi}_s\psi_s\rangle \\
M_s=m_s-G \langle \overline{\psi}_s\psi_s\rangle-\kappa \langle \overline{\psi}_u\psi_u\rangle\langle \overline{\psi}_d\psi_d\rangle
\end{array}
\right.
\end{align}
where the condensates are
\begin{align}
\label{eqcondensate}
\langle\overline{\psi}_f \psi_f\rangle=-4 M_f \int\frac{\mathrm{d}^4p}{\left(2\pi\right)^4}\frac{1}{p_4^2+p^2+M_f^2} 
\end{align}

\subsection{Inclusion of temperature, chemical potential and background magnetic field} 

The inclusion of the effect of a finite magnetic field at a Lagrangian level is done by replacing the Lagrangian (\ref{lag}) by
\begin{align}
 \mathcal{L}=\mathcal{L}_{\text{NJL}}-\dfrac{1}{4}F_{\mu\nu}F^{\mu\nu}
\end{align}
where $F_{\mu\nu}$ is the electromagnetic field tensor.
The coupling between the magnetic field $B$ and the quarks is now inside the covariant derivative $D^{\mu}=\partial^{\mu}-iq_fA^\mu_{EM}$, where $q_f$ is the quark electric charge, $A^{EM}_\mu=\delta_{\mu 2}x_1B$ is a constant magnetic field, pointing in the $z$ direction and $F_{\mu\nu}=\partial_\mu A^{EM}_{\nu}-\partial_\nu A^{EM}_\mu$. 

At the mean field level, the extension to take into account the medium effects of finite temperature and/or chemical potential can be done in the usual way by replacing the $p_4$ integration by a summation over Matsubara frequencies,
\begin{align}
p_4 &\rightarrow \pi  T (2 n + 1) - i \mu  \nonumber\\
\int \mathrm{d}p_4 &\rightarrow  2\pi T \sum _{n=-\infty }^{+\infty }\,.
\end{align}
The inclusion of the effect of a finite magnetic field can be viewed as the substitution of the integration over transverse momentum, with respect to the direction of the magnetic field by a summation over Landau levels  (denoted by the index $m$) averaged over the spin  related index, $s$,
\begin{align}
  \int\frac{\mathrm{d}^2p_\perp}{\left(2\pi\right)^2}&
  \rightarrow \frac{2\pi\left|q\right|B}{\left(2\pi\right)^2} \frac{1}{2}\sum_{s=-1,+1}\sum_{m=0}^{+\infty},\nonumber\\
  \qquad p^2_\perp &
  \rightarrow (2 m+1-s)\left|q\right| B.
\end{align}
Here we have taken the direction of z-axis as to coincide with that of the magnetic field such that $\overrightarrow{B}=B \hat{z}$.

The medium part as well as the nonmagnetic field dependent at a vanishing chemical potential (the standard vacuum part) are regularized using a three-dimensional cutoff on the spatial part of the momentum integrals whereas the magnetic field dependent contribution at a vanishing temperature and chemical potential is done as in \cite{Menezes:2008qt,Menezes:2009uc,Avancini:2011zz}. This is achieved by first performing the full sum over the Landau levels in the vacuum part, which enables the separation of the magnetic field dependent contribution from the standard, nonmagnetic field dependent, contribution. The former is then evaluated using dimensional regularization as in \cite{Menezes:2008qt}.
 
\section{Fitting LQCD results}\label{sec3}

\begin{table*}
\caption{Dynamical mass of the light sector quarks ($M_f$) as a function of the magnetic field strength ($B$) \cite{Endrodi:2019whh} as well as the errors in their estimations ($\sigma_{M_f}$).} 
\label{lQCDdata}
\begin{footnotesize}
\begin{tabular*}{0.66\textwidth}{@{\extracolsep{\fill}}c|rr|rr|rr@{}}\hline
\multicolumn{1}{c}{$eB$ [$\mathrm{GeV}^2$]} & \multicolumn{1}{c}{$M_u$ [GeV]} & \multicolumn{1}{c}{$\sigma_{M_u}$ [MeV]} & \multicolumn{1}{c}{$M_d$ [GeV]} &\multicolumn{1}{c}{$\sigma_{M_d}$ [MeV]} & \multicolumn{1}{c}{$M_s$ [GeV]} & \multicolumn{1}{c}{$\sigma_{M_s}$ [MeV]} \\
\hline
0.0 & 0.3115047 & 8.900894 & 0.3116843 & 8.852091 & 0.5500066 & 15.88578\\
0.1 & 0.3272839 & 17.06525 & 0.2933519 & 14.70449 & 0.5246419 & 23.01390\\
0.2 & 0.3341793 & 20.73042 & 0.2799108 & 17.68701 & 0.4999245 & 27.28394\\
0.3 & 0.3348220 & 19.81968 & 0.2695186 & 17.65552 & 0.4776900 & 26.89854\\
0.4 & 0.3301029 & 18.31746 & 0.2611673 & 17.98711 & 0.4568607 & 28.97203\\
0.5 & 0.3194990 & 20.62144 & 0.2581306 & 19.89098 & 0.4369001 & 34.67607\\
0.6 & 0.3037266 & 28.12296 & 0.2595464 & 24.61544 & 0.4163851 & 40.79873\\
0.7 & 0.2859020 & 34.44003 & 0.2607701 & 29.56910 & 0.4029641 & 44.11870\\
\hline
\end{tabular*}
\end{footnotesize} 
\end{table*}

As stated previously, the main purpose of this paper is to explore the consequences of the magnetic field dependence of the dynamical masses of the light sector quarks [$M_f(B), ~f\in\left\{u,d,s\right\}$] as reported in \cite{Endrodi:2019whh}  (at a vanishing temperature and chemical potential) in the framework of the NJL model (for convenience these values are listed in Table \ref{lQCDdata}). We chose to split this procedure in two steps:
\begin{enumerate}
\item fix the values of $G$, $\kappa$ and $\Lambda$ and the current masses ($m_f, ~f\in\left\{u,d,s\right\}$) at a vanishing magnetic field strength;
\item use the values of two of the dynamical quark masses at finite magnetic field strength to fit the coupling strengths of the interactions thus introducing a magnetic field strength dependence on them [$G\left(B\right)$ and $\kappa\left(B\right)$)] while keeping $\Lambda$ and $m_f$ fixed.
\end{enumerate}

\subsection{Fits at vanishing magnetic field}

There are 6 degrees of freedom in our fit: current masses, coupling strengths and cutoff. Three of these should be fixed using the dynamical masses (through Eqs. \ref{GapEqs}) as the main idea behind this paper relies on taking at face value the LQCD values for $M_f$. Let us think of the coupling strengths and cutoff as being fixed by these conditions. That leaves us with the choice of the three current masses. 

\begin{figure*}
\center
\subfigure[]{\label{grafGMpi0140MK0494msvar}\includegraphics[width=0.32\textwidth]{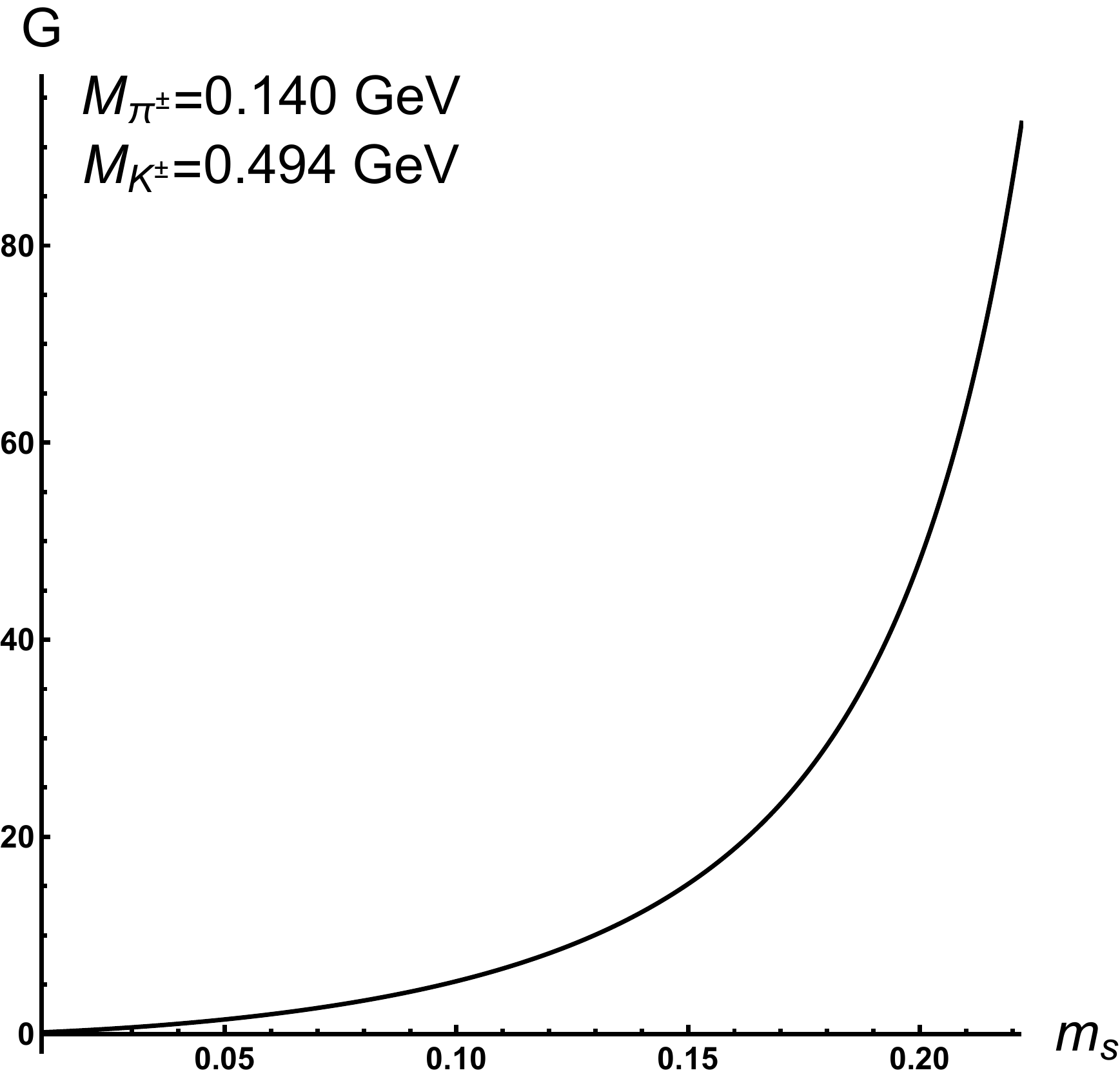}}
\subfigure[]{\label{grafKMpi0140MK0494msvar}\includegraphics[width=0.32\textwidth]{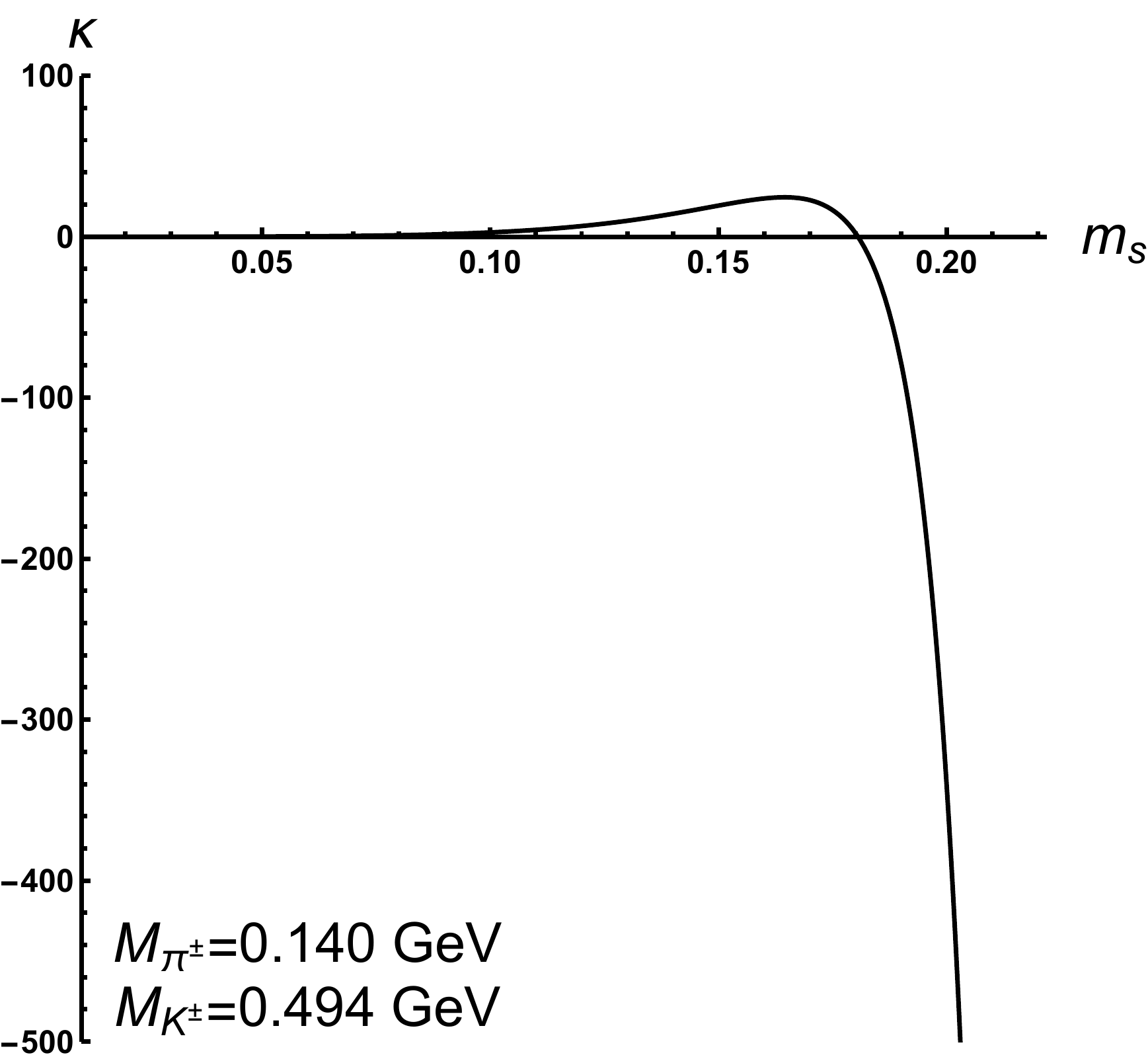}}
\subfigure[]{\label{grafLambdaMpi0140MK0494msvar}\includegraphics[width=0.32\textwidth]{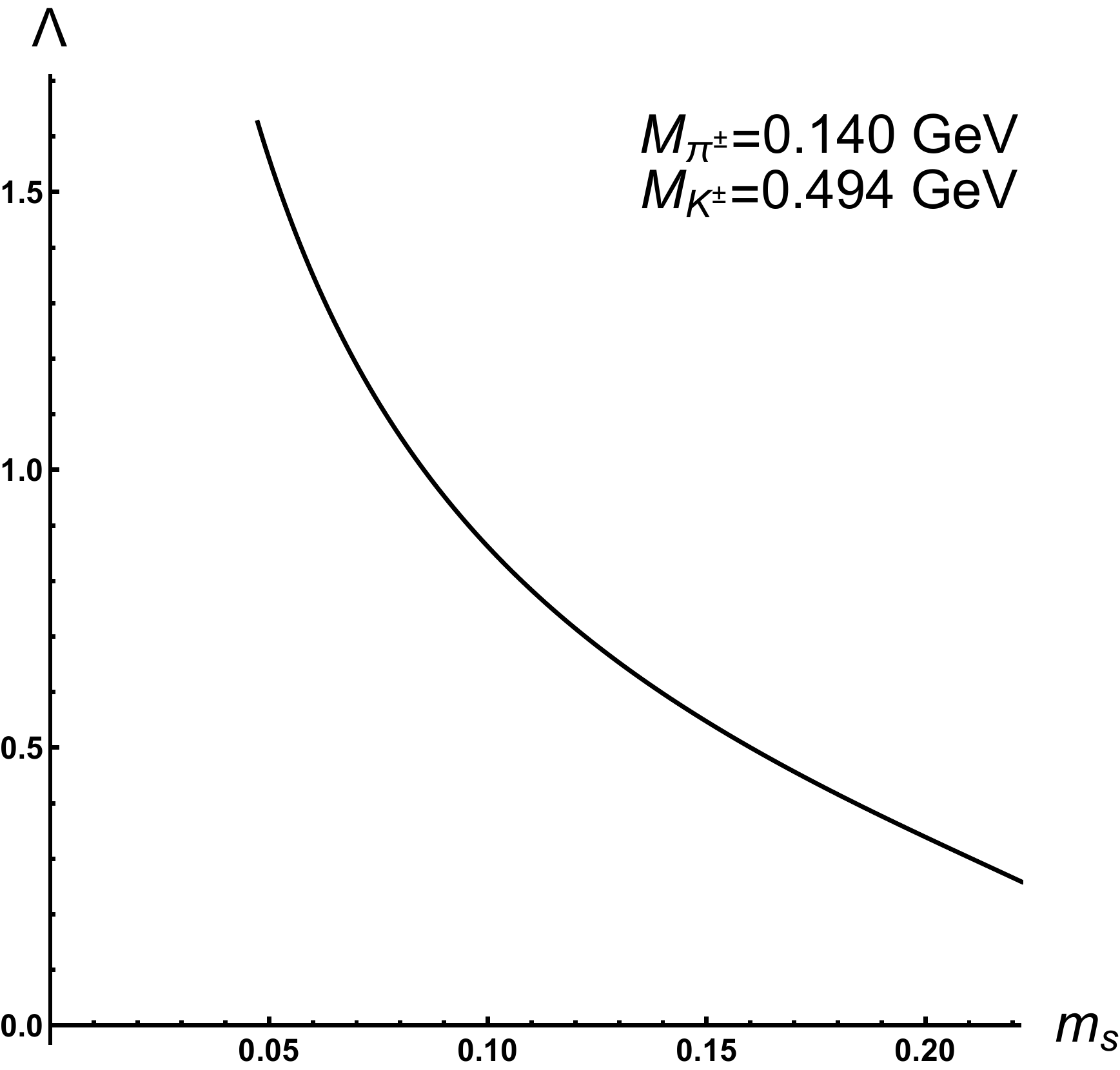}}
\caption{
Coupling strengths and cutoff dependence on the choice of a strange quark current mass and a fit to reproduce the physical masses of the charged pion and kaon.
}
\label{FitMpi0140MK0494msvar}
\end{figure*}

Naively one could expect to be able to fit these using the mass of some of the lightest pseudoscalar mesons. As the NJL model, by construction, relies on the relevance of chiral and axial symmetry breaking (spontaneous and explicit in the case of the former and explicit, through the 't Hooft determinant, in the case of the latter) using the pion, kaon and eta prime meson masses seems the obvious choice. Some other reasonability criteria, such as the size of the cutoff and the sign of the coupling strengths (or more precisely its consequences in the obtained meson spectra), will however come into play.

Let us start by fixing the charged pion and kaon to their physical values fitting $m_u$ and $m_d$ and leaving $m_s$ as a free parameter. In Figs. \ref{FitMpi0140MK0494msvar} the results of the fit are presented. A negative coupling constant for the 't Hooft determinant is only obtained for strange quark current mass above a critical value of $m_s>0.180~\mathrm{GeV}$.

\begin{figure*}
\center
\subfigure[]{\label{grafMpsMesonsMpi0140MK0494msvar}\includegraphics[width=0.32\textwidth]{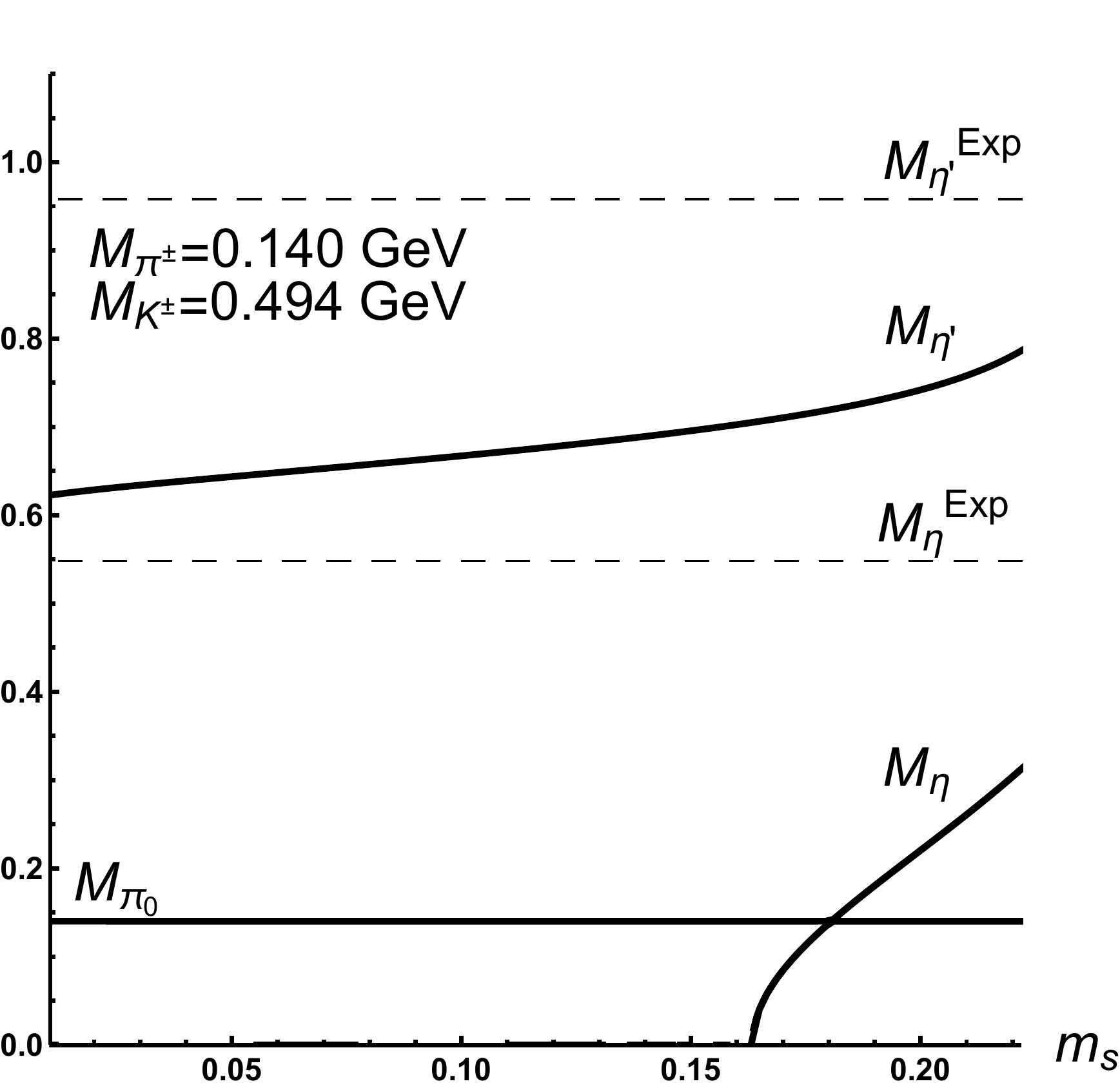}}
\subfigure[]{\label{grafGammaEtaMesonsMpi0140MK0494msvar}\includegraphics[width=0.32\textwidth]{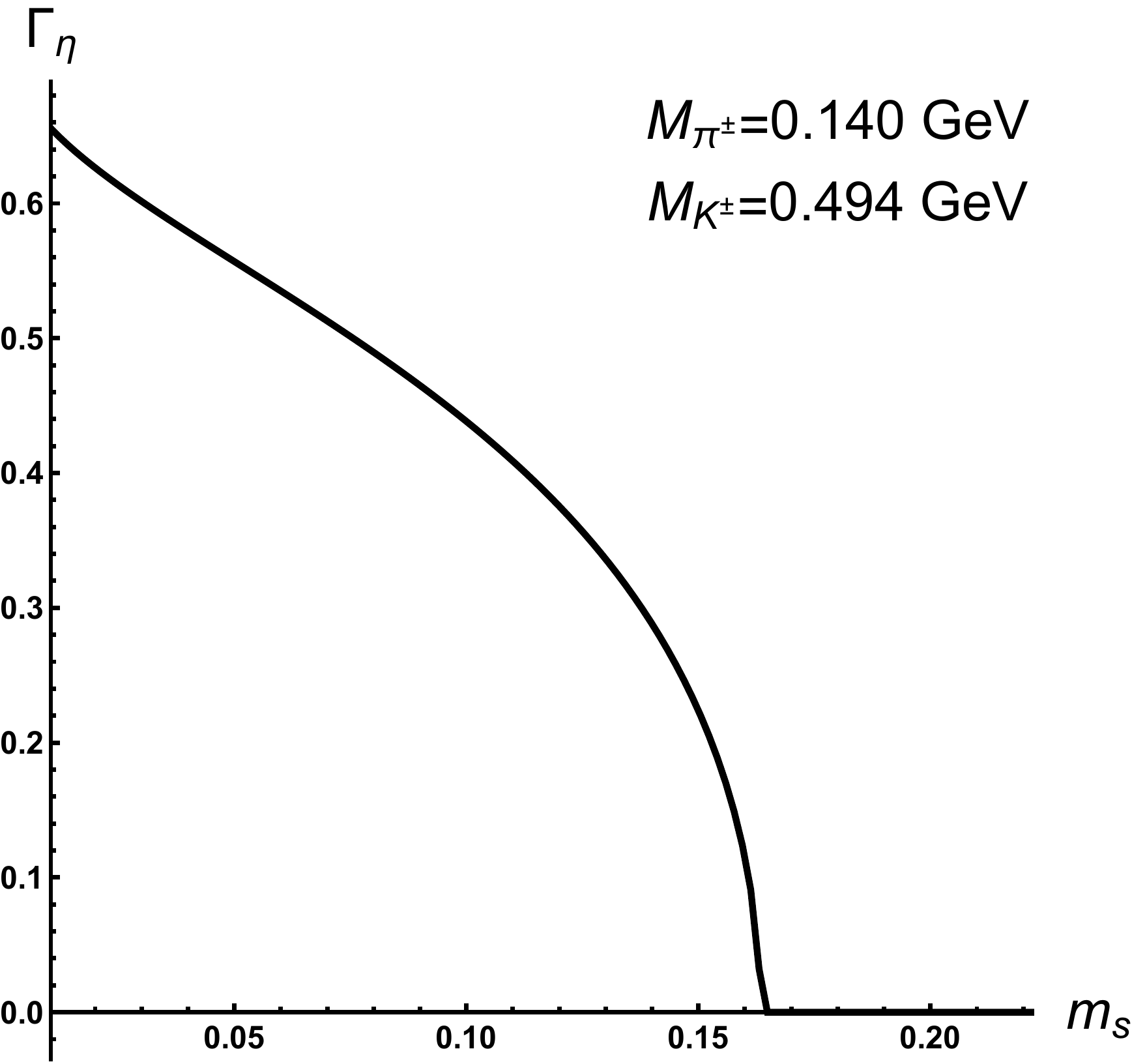}}
\subfigure[]{\label{grafGammaEtaPrimeMesonsMpi0140MK0494msvar}\includegraphics[width=0.32\textwidth]{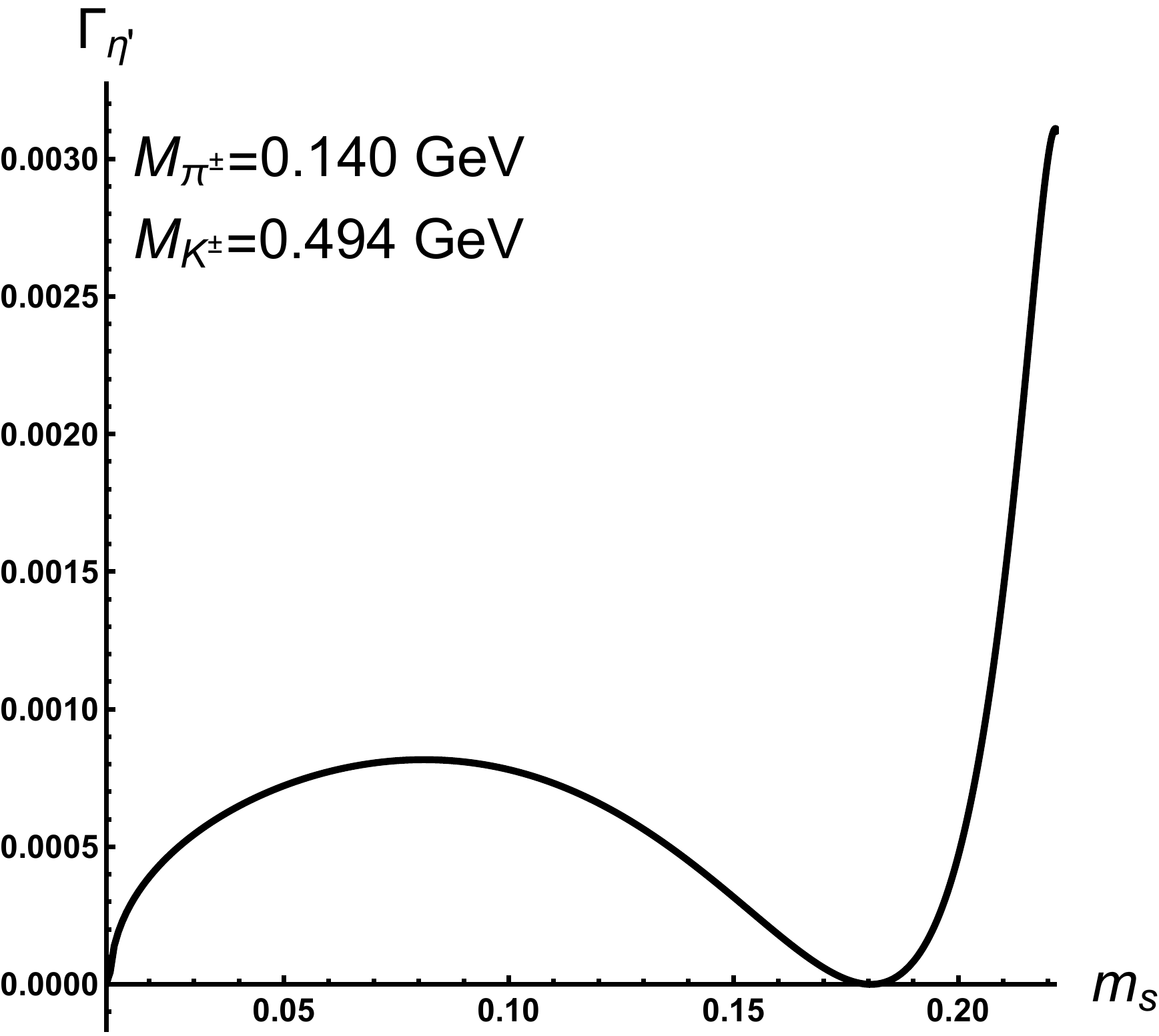}}
\caption{Pseudoscalar meson masses dependence on the choice of strange quark current mass in Fig. \ref{grafMpsMesonsMpi0140MK0494msvar}. The lightest solution drops below that of the pion for $\kappa>0$ (which happens for this fit for $m_s<0.180~\mathrm{GeV}$) and goes to zero below the critical value of $m_s=0.163~\mathrm{GeV}$. The imaginary part of the pole of the propagator (for a given meson $X$ the pole is located at $M_{X}-\frac{\imath}{2}\Gamma_X$, with $\Gamma_X$ corresponding to the decay width) is presented for the $\eta$ meson in Fig. \ref{grafGammaEtaMesonsMpi0140MK0494msvar} (finite for the massless solution) and for the $\eta'$ in Fig. \ref{grafGammaEtaPrimeMesonsMpi0140MK0494msvar}. It is noteworthy that the decay with for the $\eta'$ vanishes for $\kappa=0$, when it is degenerate with $\pi_0$.
}
\label{FitMpi0140MK0494msvarII}
\end{figure*}

\begin{figure*}
\center
\subfigure[]{\label{grafLambdavsKMpi0140MK0494msvar}\includegraphics[width=0.32\textwidth]{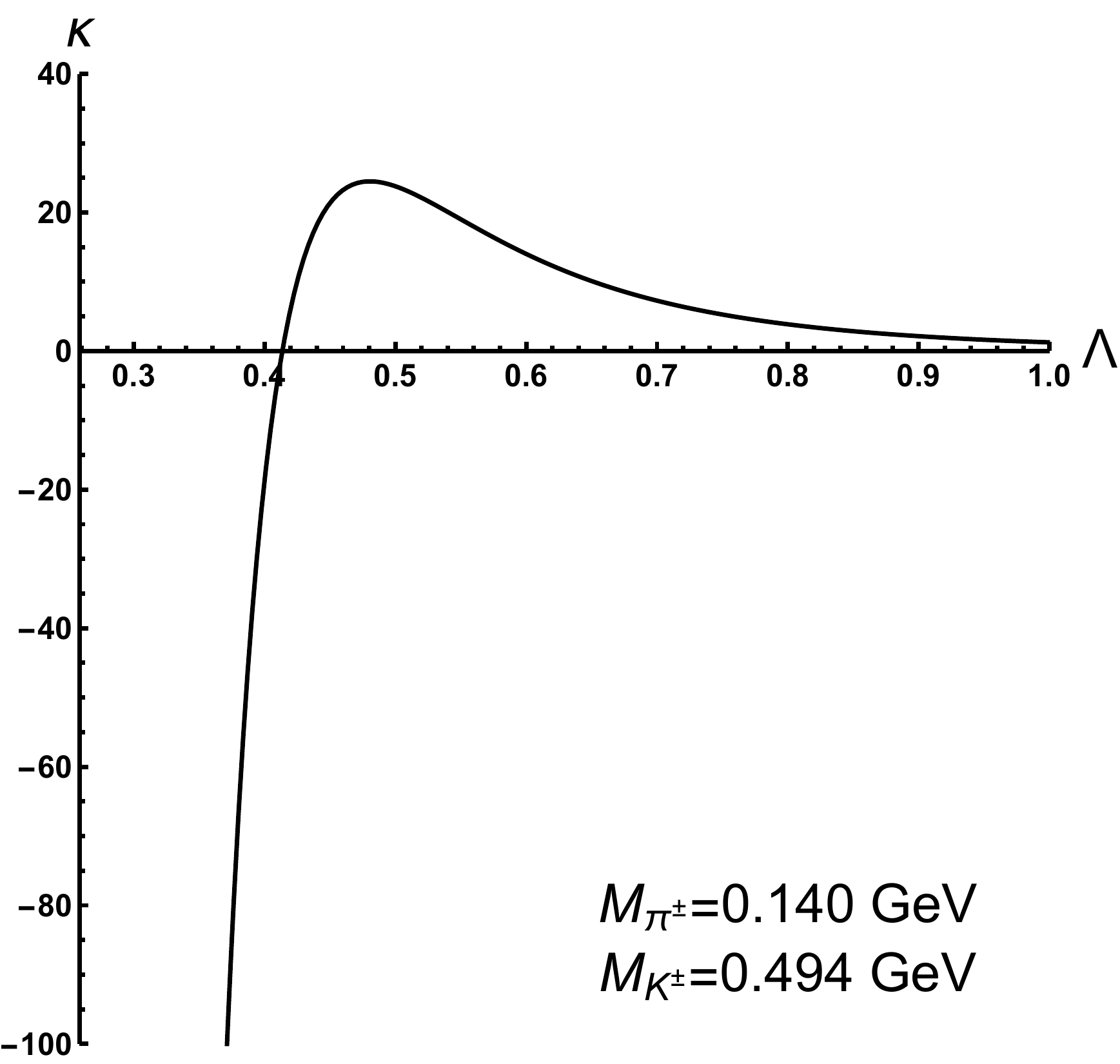}}
\caption{Dependence of the coupling strength of the 't determinant interaction on the cutoff imposing both the pion and kaon masses on the physical values.}
\label{NoPhysicalFit}
\end{figure*}

As can be seen in Figs. \ref{grafMpsMesonsMpi0140MK0494msvar} at a positive $\kappa$ ($m_s<0.180~\mathrm{GeV}$) one of the neutral light pseudoscalars becomes lighter than the pion. Below a critical value of the strange quark current mass ($m_s<0.163~\mathrm{GeV}$) it becomes massless and, as can be seen in Fig. \ref{grafGammaEtaMesonsMpi0140MK0494msvar}, gains a finite decay width. As can be seen in Fig. \ref{grafGammaEtaPrimeMesonsMpi0140MK0494msvar}, the decay width of the $\eta'$ meson vanishes for the choice of $m_s$ corresponding to vanishing $\kappa$ [see Fig. \ref{grafKMpi0140MK0494msvar}] which also results in degenerate $\pi_0$ and $\eta$. These positive $\kappa$ scenarios are unphysical and avoiding them results in a maximum value of the cutoff $\Lambda < 0.414~\mathrm{GeV}$ as can be seen in Fig. \ref{NoPhysicalFit}.

\begin{figure*}
\center
\subfigure[]{\label{grafLambdaMpi0140MEtaPrime0958MKpmvar}\includegraphics[width=0.32\textwidth]{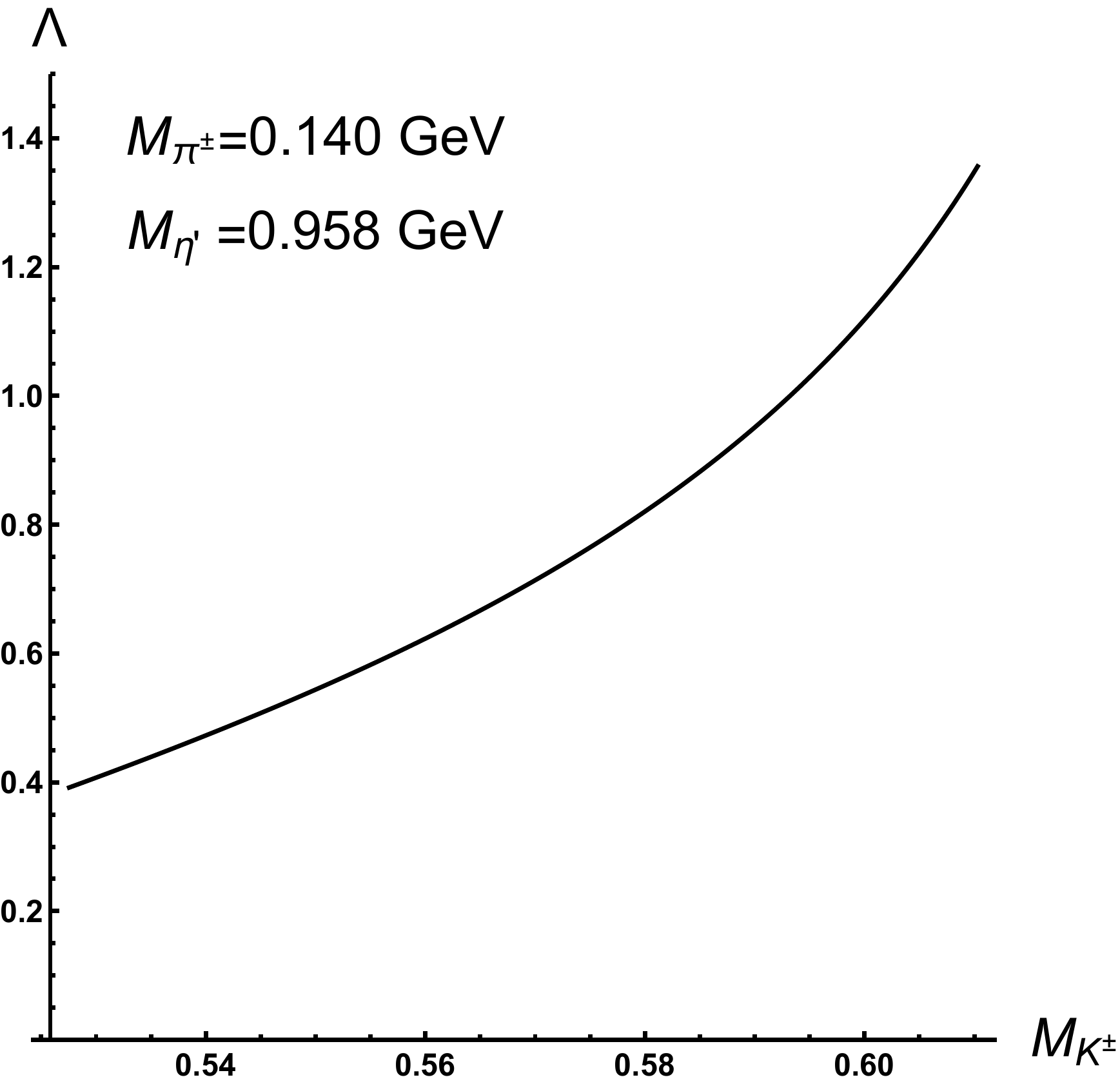}}
\subfigure[]{\label{grafMEtaMpi0140MEtaPrime0958MKpmvar}\includegraphics[width=0.32\textwidth]{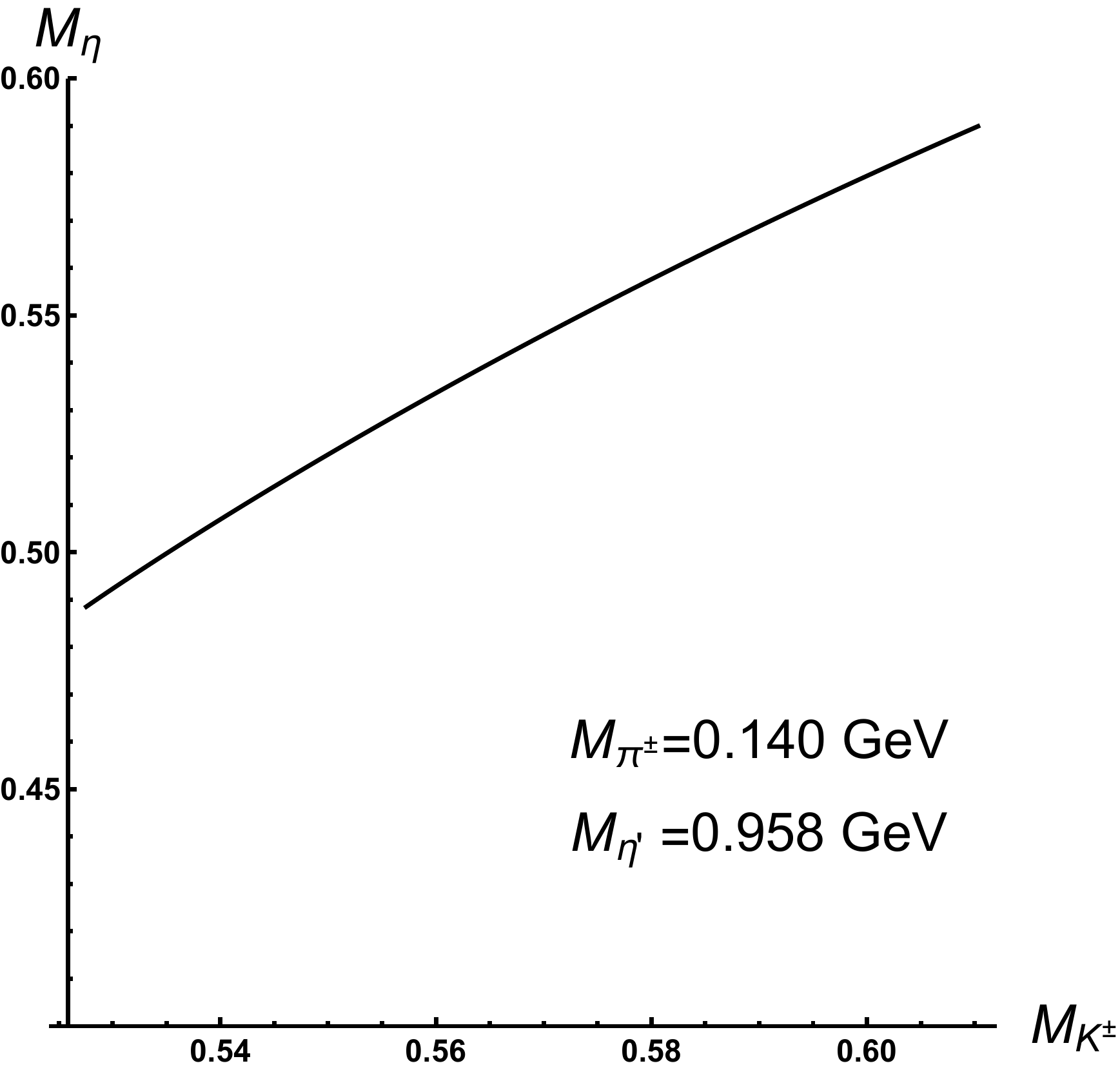}}
\subfigure[]{\label{grafGammaEtaPrimeMpi0140MEtaPrime0958MKpmvar}\includegraphics[width=0.32\textwidth]{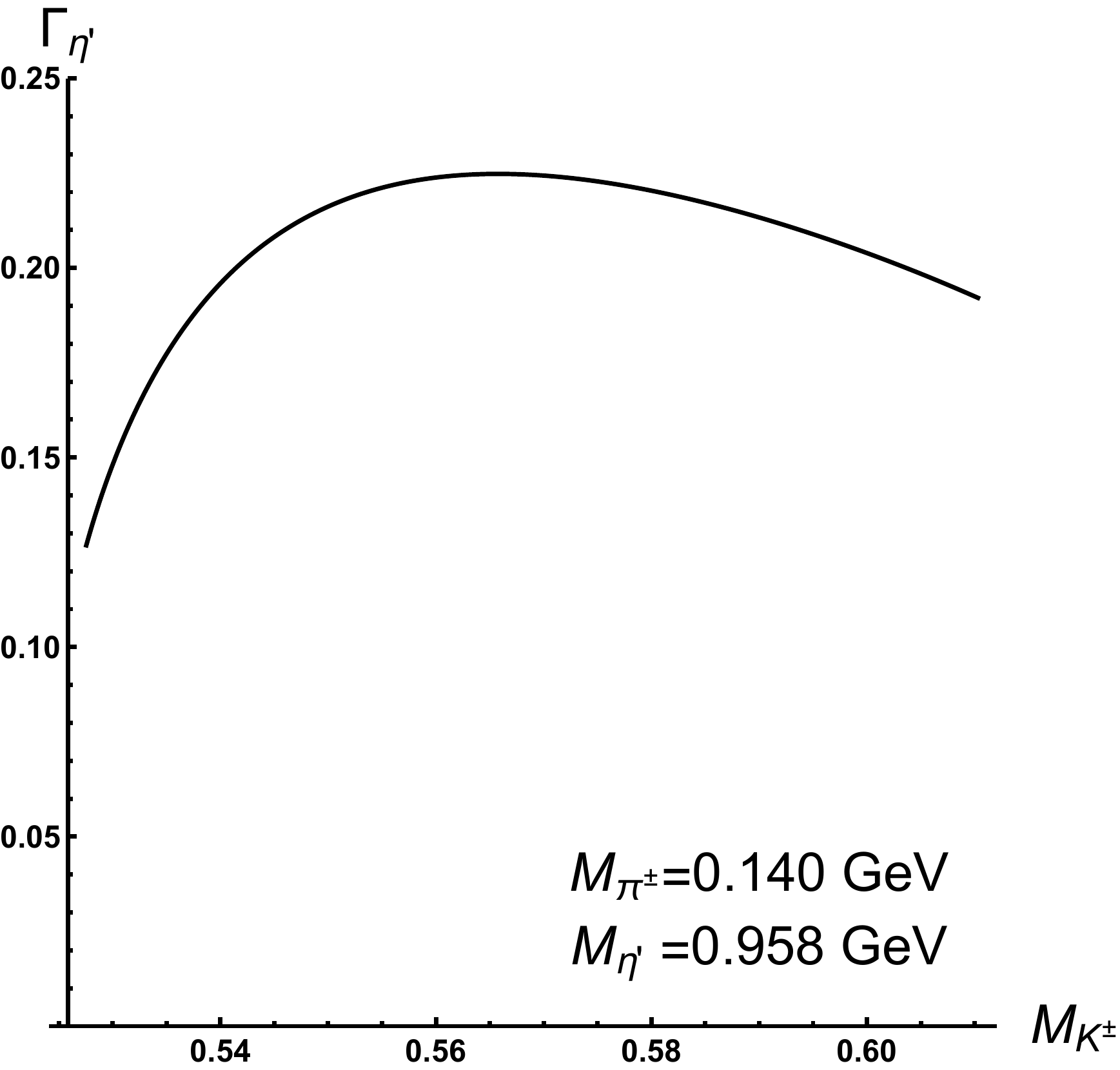}}
\caption{From left to right the dependence of the cutoff, mass of the eta meson and decay width of the eta prime on the choice of kaon mass while keeping the pion and eta prime meson masses fixed to their experimental values (all in $\mathrm {GeV}$).}
\label{CutOffFits}
\end{figure*}

A simultaneous fit of $M_\pi$, $M_K$ and $M_{\eta'}$ could not be achieved in the preformed scan [which already goes into unreasonably low values of the cutoff as can be seen in Fig. \ref{grafLambdaMpi0140MK0494msvar}]. It should be noted, however, that raising the value $M_K$ enables us to reach the physical value of $M_{\eta'}$. In Fig. \ref{CutOffFits} one can see the dependence of the cutoff, eta meson mass and eta prime decay width on the choice of mass for the kaon, while keeping the pion and eta prime meson masses at their experimental values. In Tables \ref{ParameterSetsI} and \ref{ParameterSetsISpectra} some parameter sets resulting in cutoffs of $0.500$, $0.600$, $0.700~\mathrm{GeV}$ are presented as well as a set which reproduces the physical values of the masses of pion, eta prime and eta mesons.

\begin{table*}
\caption{Parameter sets obtained fitting the dynamical masses of the quarks to the values reported in \cite{Endrodi:2019whh}, the mass of the pion ($M_{\pi^\pm}=0.140~\mathrm{GeV}$) and the eta prime mesons ($M_{\eta}=0.958~\mathrm{GeV}$) at a vanishing magnetic field strength. Sets $a$, $b$ and $c$ were chosen so as to have a cutoff of $0.500$, $0.600$ and $0.700~\mathrm{GeV}$ respectively. Set $d$ is chosen so as to reproduce the eta meson mass $M_\eta=0.548~\mathrm{GeV}$.} 
\label{ParameterSetsI}
\begin{footnotesize}
\begin{tabular*}{0.66\textwidth}{@{\extracolsep{\fill}}l|rrr|rr|r@{}}\hline
\multicolumn{1}{c}{} &
\multicolumn{1}{c}{$m_u$ [MeV]} & \multicolumn{1}{c}{$m_d$ [MeV]} & \multicolumn{1}{c}{$m_s$ [MeV]} 
&\multicolumn{1}{c}{$G~\left[\text{GeV}^{-2}\right]$} & \multicolumn{1}{c}{$\kappa~\left[\text{GeV}^{-5}\right]$} & \multicolumn{1}{c}{$\Lambda~\left[\text{GeV}\right]$} \\
\hline
a) &7.06242 & 7.19017 & 196.403 & 14.2933  & -409.947 & 0.500000 \\
b) &5.89826 & 6.01442 & 180.450 &  9.47574 & -149.340 & 0.600000 \\
c) &4.99322 & 5.09877 & 165.328 &  6.77511 & -63.7862 & 0.700000 \\
d) &4.75221 & 4.85464 & 160.820 &  6.17046 & -50.1181 & 0.731313\\
\hline
\end{tabular*}
\end{footnotesize} 
\end{table*}

\begin{table*}
\caption{Kaon and eta meson masses along with the eta prime decay width for the sets listed in Table \ref{ParameterSetsI}.} 
\label{ParameterSetsISpectra}
\begin{footnotesize}
\begin{tabular*}{0.66\textwidth}{@{\extracolsep{\fill}}l|rrr@{}}\hline
\multicolumn{1}{c}{} &
\multicolumn{1}{c}{$M_{K^\pm}$ [GeV]} & \multicolumn{1}{c}{$\Gamma_{\eta'}$ [MeV]} & \multicolumn{1}{c}{$M_\eta$ [GeV]}\\
\hline
a) & 0.543936 & 0.205968 & 0.512422 \\
b) & 0.557196 & 0.222571 & 0.530023\\
c) & 0.568573 & 0.224593 & 0.544170\\
d) & 0.571768 & 0.223918 & 0.548000\\
\hline
\end{tabular*}
\end{footnotesize} 
\end{table*}

\subsection{Fitting the magnetic field dependence}

For the second point, pertaining to the magnetic field dependence, we kept the cutoff ($\Lambda$) and the current masses ($m_f$) fixed while fitting the couplings strengths, which thus gain a magnetic field dependence  [$G(B)$ and $\kappa(B)$], using two of the dynamical masses. The results presented here were obtained by fitting the \emph{down} and \emph{strange} quarks dynamical mass. The magnetic field dependence of the dynamical mass of the \emph{up} quark can therefore be used as one of the criteria to check the adequacy of the several scenarios. An attempt to use the mass of the up and down quarks as inputs in this fit (leaving the dynamical mass of the strange quark as output) was not successful.

\begin{figure*}
\center
\subfigure[]{\label{grafMilQCDMPi0140MEtal0958Lambda050006000700MEta0548}\includegraphics[width=0.32\textwidth]
{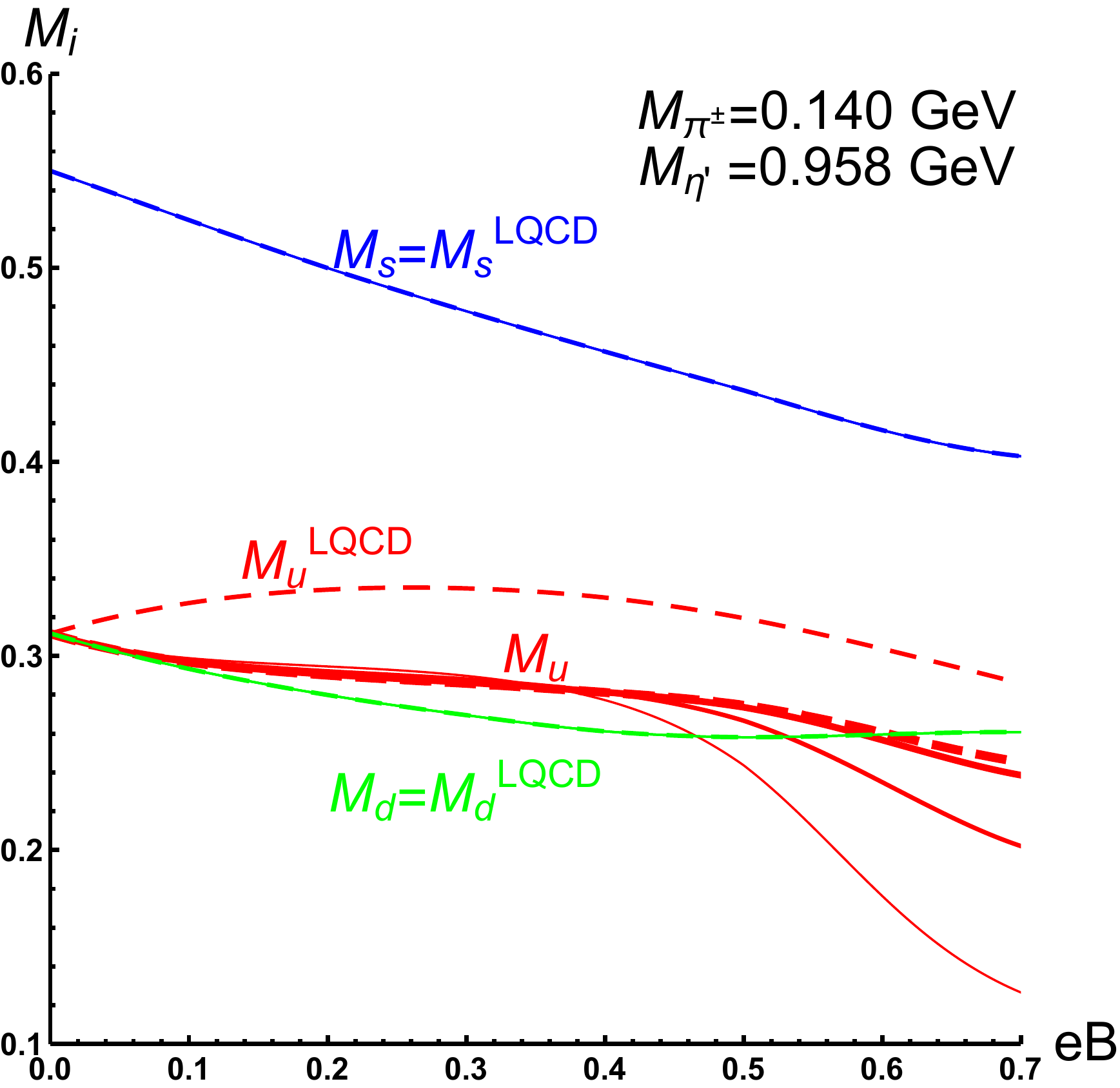}}
\subfigure[]{\label{grafGvsBlQCDMPi0140MEtal0958Lambda050006000700MEta0548}\includegraphics[width=0.32\textwidth]
{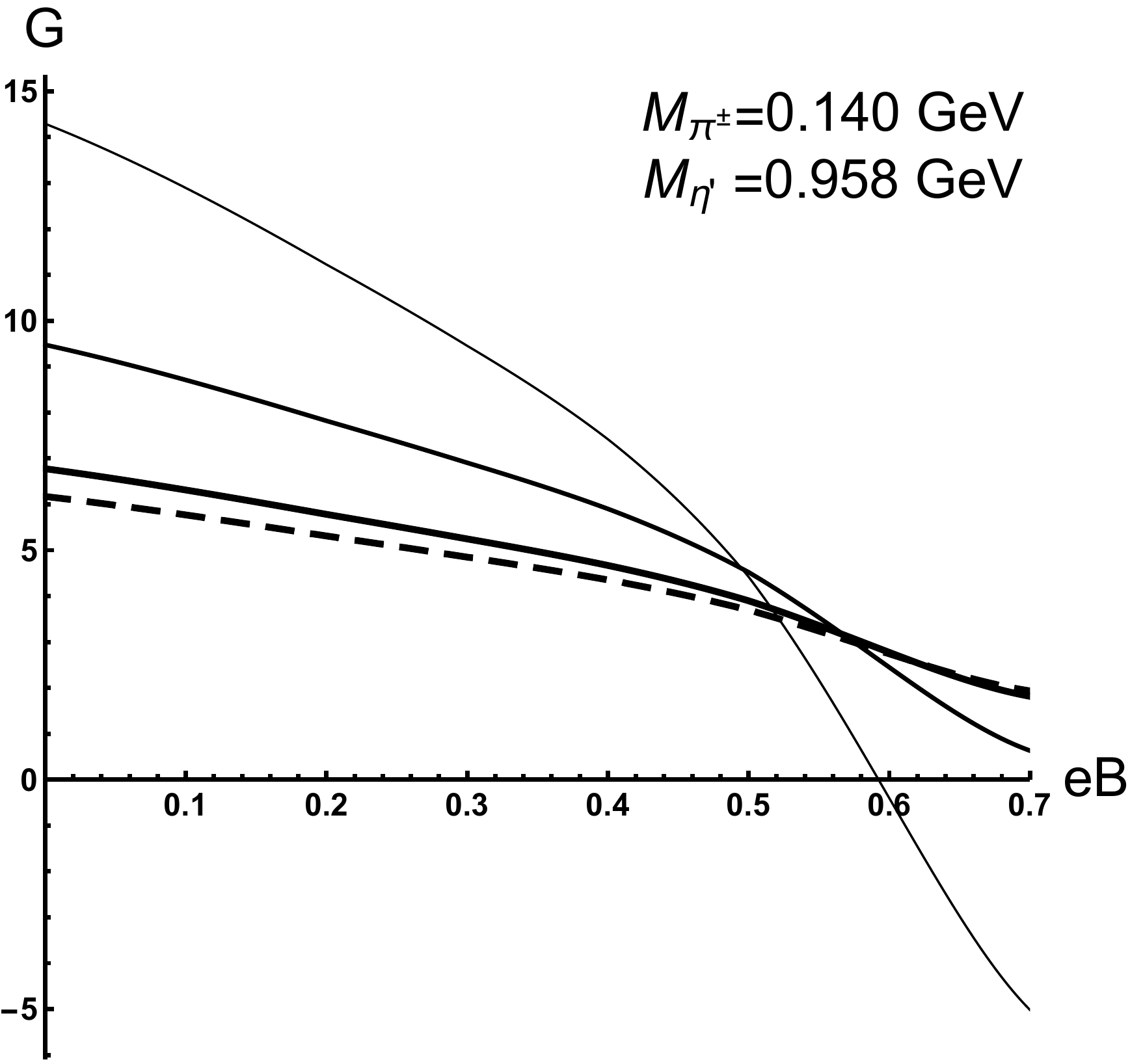}}
\subfigure[]{\label{grafKvsBlQCDMPi0140MEtal0958Lambda050006000700MEta0548}\includegraphics[width=0.32\textwidth]
{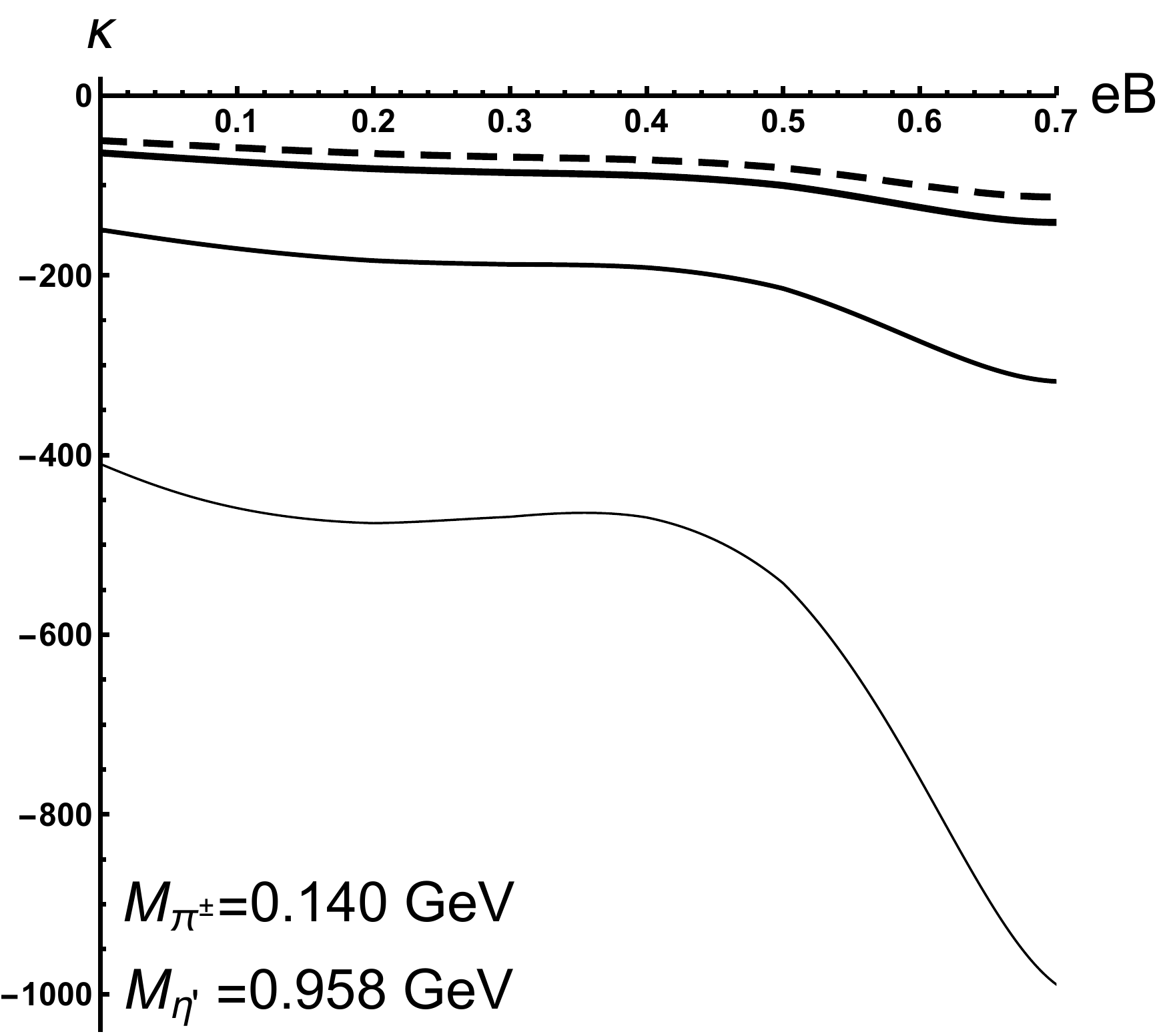}}
\caption{Results of the magnetic field strength dependent fit of the couplings: $G(B)$ and $\kappa(B)$. In \ref{grafMilQCDMPi0140MEtal0958Lambda050006000700MEta0548} the magnetic field dependence of the dynamical masses of the quarks \cite{Endrodi:2019whh}. The fit is done by imposing the reproduction of the LQCD results for $M_d$ and $M_s$, as such, $M_u$ is an output. In increasing thickness the full lines labeled $M_u$ correspond to $\Lambda=0.500~\mathrm{GeV}$, $0.600~\mathrm{GeV}$ and $0.700~\mathrm{GeV}$ (sets $a$, $b$ and $c$ in Table \ref{ParameterSetsI}) whereas the thick dashed line corresponds to the choice of cutoff which reproduces the eta meson mass $M_\eta=0.548~\mathrm{GeV}$ (set $d$ in Table \ref{ParameterSetsI}). All sets reproduce the pion and the eta prime masses.}
\label{ParamsBdep}
\end{figure*}

The results of the magnetic field dependent fit of the couplings can be seen in the Figs. \ref{ParamsBdep} and Table \ref{MagDepParamsNegKappa}. The split between $M_u$ and $M_d$ is smaller for our parameter sets [see Fig. \ref{grafMilQCDMPi0140MEtal0958Lambda050006000700MEta0548}] and starts to deviate markedly for larger fields. This larger deviation appears however to be related to regularization effects as it occurs for larger magnetic field strengths when we move to higher cutoffs. These regularization effects are also patent when considering the dependence of the couplings [see Figs. \ref{grafGvsBlQCDMPi0140MEtal0958Lambda050006000700MEta0548} and \ref{grafKvsBlQCDMPi0140MEtal0958Lambda050006000700MEta0548}]. There is a clear onset of a deviation from the behavior that is observed at lower magnetic field strengths which occurs at smaller field strengths for smaller cutoff. For the sets with larger cutoff (sets $c$ and $d$) the behavior is approximately linear with a decreasing positive $G$ as a function of the magnetic field and an increasing contribution coming from the 't Hooft determinant interaction as we see a negative $\kappa$ increasing in absolute value. This increase in the 't Hooft term relevance is possibly connected to an increase in the relevance of $U_A(1)$ due to the quark spin interaction with electromagnetic fields \cite{Guo:2017dzf}.

\begin{table*}
\caption{Values of the coupling strengths of the model at sample values for the magnetic field fitted as to reproduce the dynamical masses of the down and strange quarks (as reported in \cite{Endrodi:2019whh}) and using the current quark masses and regularization cutoff from Table \ref{ParameterSetsI}.} 
\label{MagDepParamsNegKappa}
\begin{footnotesize}
\begin{tabular*}{0.85\textwidth}{@{\extracolsep{\fill}}l|rr|rr|rr|rr@{}}\hline
\multicolumn{1}{c}{}  &
\multicolumn{2}{c}{$a)$} &
\multicolumn{2}{c}{$b)$} &
\multicolumn{2}{c}{$c)$} &
\multicolumn{2}{c}{$d)$}
\\
\multicolumn{1}{c}{$eB$}  &
\multicolumn{1}{c}{$G$} &
\multicolumn{1}{c}{$\kappa$} &
\multicolumn{1}{c}{$G$} &
\multicolumn{1}{c}{$\kappa$} &
\multicolumn{1}{c}{$G$} &
\multicolumn{1}{c}{$\kappa$} &
\multicolumn{1}{c}{$G$} &
\multicolumn{1}{c}{$\kappa$} 
\\
\multicolumn{1}{c}{$\left[\text{GeV}\right]$}  &
\multicolumn{1}{c}{$\left[\text{GeV}^{-2}\right]$} &
\multicolumn{1}{c}{$\left[\text{GeV}^{-5}\right]$} &
\multicolumn{1}{c}{$\left[\text{GeV}^{-2}\right]$} &
\multicolumn{1}{c}{$\left[\text{GeV}^{-5}\right]$} &
\multicolumn{1}{c}{$\left[\text{GeV}^{-2}\right]$} &
\multicolumn{1}{c}{$\left[\text{GeV}^{-5}\right]$} &
\multicolumn{1}{c}{$\left[\text{GeV}^{-2}\right]$} &
\multicolumn{1}{c}{$\left[\text{GeV}^{-5}\right]$} 
 \\
\hline
0.000&  14.2933& -409.947&  9.47574& -149.34 & 6.77511& -63.7862& 6.17046& -50.1181 \\
0.025&  13.9768& -425.237&  9.30076& -155.053& 6.66863& -66.2936& 6.07794& -52.1048 \\
0.050&  13.6360& -438.726&  9.11346& -160.544& 6.55517& -68.8060& 5.97949& -54.1126 \\
0.075&  13.2740& -450.005&  8.91565& -165.649& 6.43581& -71.2560& 5.87602& -56.0895 \\
0.100&  12.8933& -459.012&  8.70884& -170.285& 6.31147& -73.6004& 5.76833& -58.0012 \\
0.125&  12.4960& -465.883&  8.49437& -174.415& 6.18300& -75.8110& 5.65714& -59.8236 \\
0.150&  12.0840& -470.836&  8.27354& -178.029& 6.05121& -77.8674& 5.54317& -61.5385 \\
0.175&  11.6593& -474.099&  8.04770& -181.123& 5.91694& -79.7513& 5.42716& -63.1290 \\
0.200&  11.2244& -475.857&  7.81835& -183.686& 5.78113& -81.4428& 5.30991& -64.5773 \\
0.225&  10.7980& -474.748&  7.59499& -185.209& 5.64925& -82.7222& 5.19612& -65.7105 \\
0.250&  10.3624& -472.860&  7.36897& -186.329& 5.51640& -83.8301& 5.08159& -66.7121 \\
0.275&  9.91500& -470.722&  7.13923& -187.186& 5.38198& -84.8125& 4.96583& -67.6159 \\
0.300&  9.45341& -468.778&  6.90474& -187.905& 5.24545& -85.7123& 4.84837& -68.4533 \\
0.325&  8.98913& -465.858&  6.67073& -188.101& 5.10956& -86.3739& 4.73150& -69.1042 \\
0.350&  8.50145& -464.395&  6.42799& -188.553& 4.96940& -87.0983& 4.61110& -69.7986 \\
0.375&  7.98110& -465.306&  6.17265& -189.561& 4.82289& -88.0050& 4.48542& -70.6284 \\
0.400&  7.41725& -469.525&  5.90025& -191.437& 4.66770& -89.2195& 4.35249& -71.6908 \\
0.425&  6.78212& -479.527&  5.59935& -194.981& 4.49788& -91.0640& 4.20734& -73.2350 \\
0.450&  6.07885& -494.515&  5.27189& -199.917& 4.31445& -93.4357& 4.05079& -75.1837 \\
0.475&  5.29586& -515.254&  4.91364& -206.442& 4.11522& -96.4163& 3.88102& -77.6013 \\
0.500&  4.41941& -542.789&  4.51957& -214.790& 3.89762& -100.101& 3.69584& -80.5623 \\
0.525&  3.34654& -585.110&  4.04499& -227.351& 3.63731& -105.472& 3.47463& -84.8363 \\
0.550&  2.16839& -636.099&  3.53317& -241.747& 3.35846& -111.524& 3.23796& -89.6340 \\
0.575& 0.907604& -695.043&  2.99558& -257.393& 3.06752& -117.988& 2.99134& -94.7446 \\
0.600&-0.403001& -760.236&  2.44764& -273.513& 2.77300& -124.521& 2.74199& -99.8967 \\
0.625& -1.71787& -828.289&  1.90952& -289.085& 2.48582& -130.687& 2.49918& -104.747 \\
0.650& -2.97697& -893.873&  1.40666& -302.800& 2.21958& -135.947& 2.27440& -108.874 \\
0.675& -4.10572& -950.085& 0.969434& -313.073& 1.99032& -139.679& 2.08117& -111.791\\
0.700& -5.01754& -989.212& 0.631820& -318.137& 1.81575& -141.211& 1.93440& -112.978 \\
\hline
\end{tabular*}
\end{footnotesize} 
\end{table*}

\begin{figure*}
\center
\subfigure[]{\label{FaixasCondensadoIMClQCD}\includegraphics[width=0.32\textwidth]{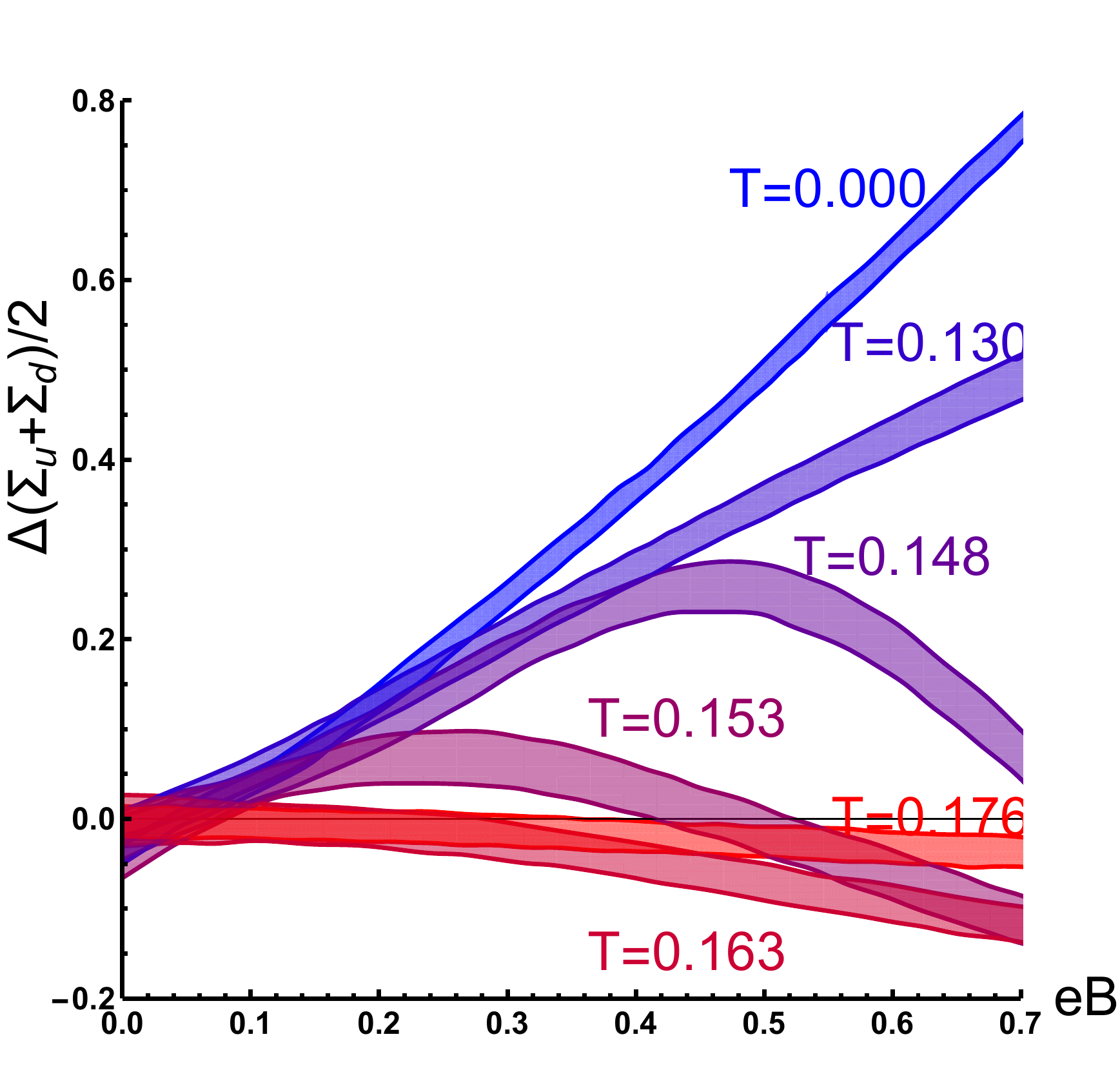}}
\caption{Renormalized chiral condensate change with magnetic field at different temperatures as given by LQCD calculations \cite{Bali:2012zg} displaying the behavior known as inverse magnetic catalysis. The chiral condensate dependence on the magnetic field strength is qualitatively different for temperatures depending on its relation to the chiral transition temperature, $T^\chi_c$. For temperatures well below $T^\chi_c$ it increases monotonically with $B$ (slope decreases with increasing temperature) whereas for temperatures well above $T^\chi_c$ it decreases monotonically (slope increases with temperature). Close to the transition temperature it exhibits a nonmonotonic behavior with a local maxima.}
\label{CondvsBIMClQCD}
\end{figure*}

Recent LQCD simulations \cite{Bali:2011qj,Bali:2012zg} have shed light into an interesting interplay of temperature and magnetic field concerning chiral symmetry. On the one hand they point to a decrease in the critical temperature for chiral (partial) restoration with increasing magnetic field strength, while on the other hand, when looking at the change in the renormalized chiral condensate due to the magnetic field, the LQCD estimates point to an increasing condensate with magnetic field strength for temperatures well below the critical temperature for chiral restoration, a decrease well above said temperature and a nonmonotonic behavior close to that temperature (an increase followed by a decrease): the inverse magnetic catalysis phenomenon.

\begin{figure*}
\center
\subfigure[]{\label{grafSigmaIMCMdMslQCDMPi0140MEtal0958Lambda0500PC}\includegraphics[width=0.32\textwidth]{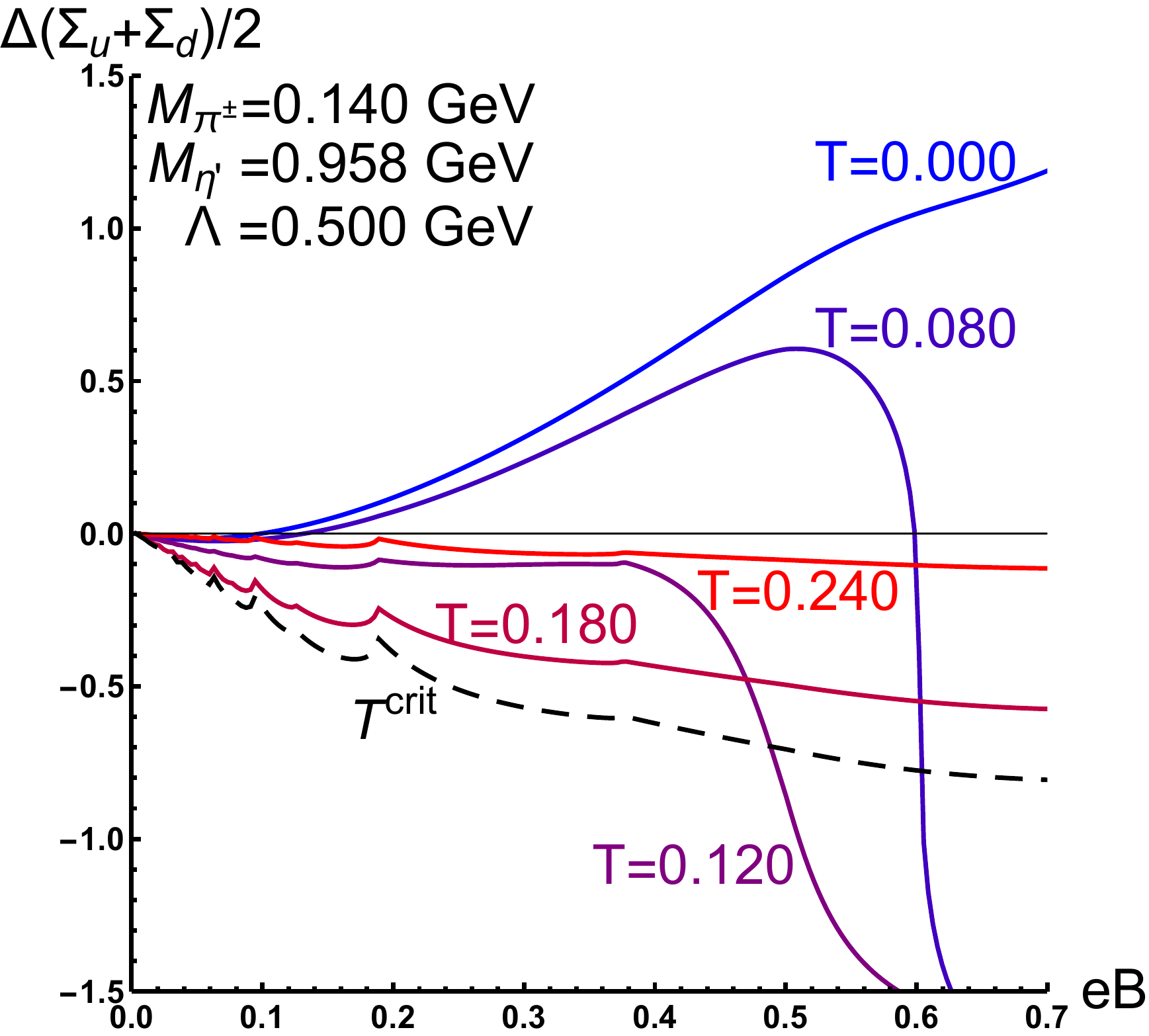}}
\subfigure[]{\label{grafSigmaIMCMdMslQCDMPi0140MEtal0958Lambda0600PC}\includegraphics[width=0.32\textwidth]{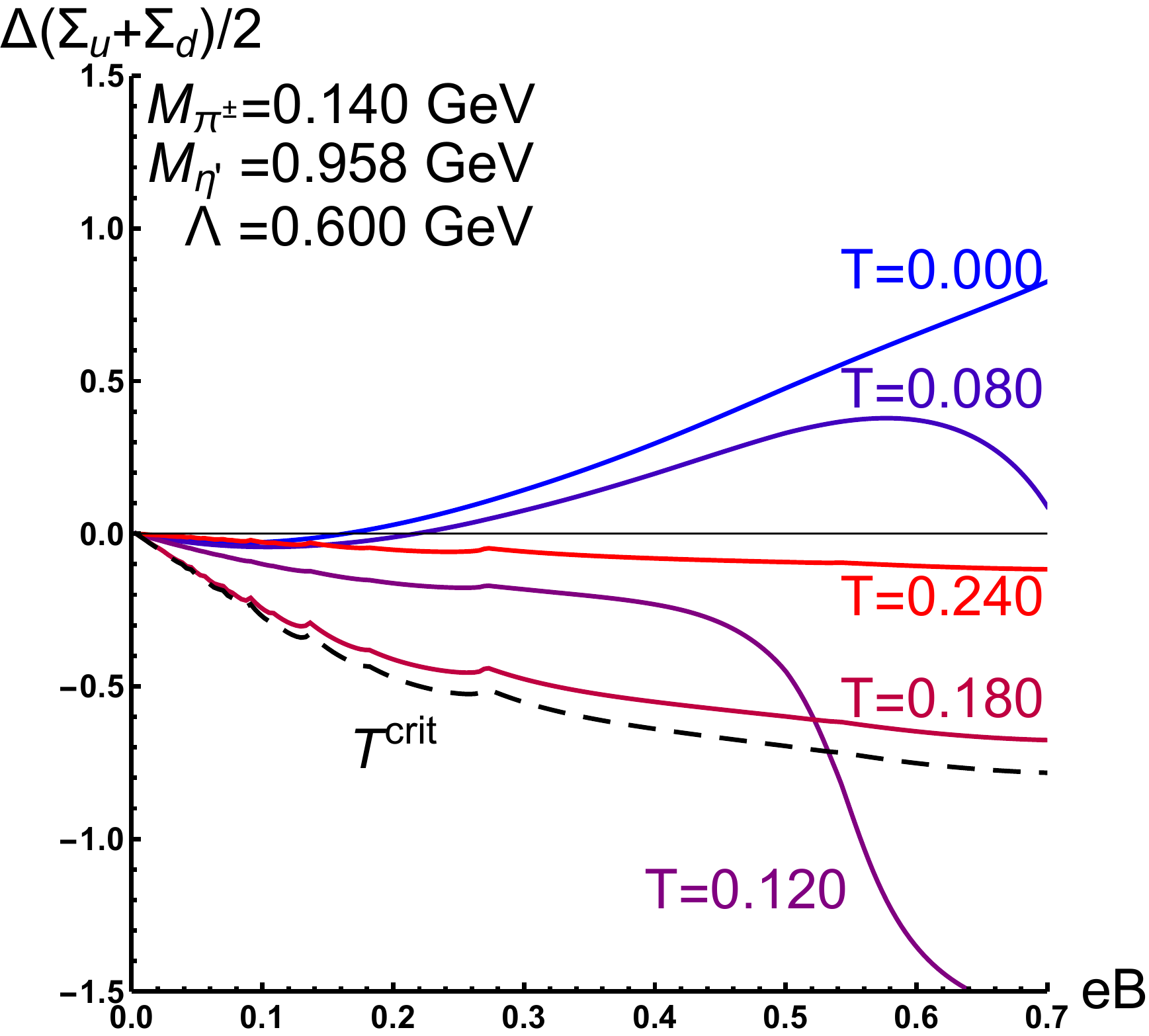}}\\
\subfigure[]{\label{grafSigmaIMCMdMslQCDMPi0140MEtal0958Lambda0700PC}\includegraphics[width=0.32\textwidth]{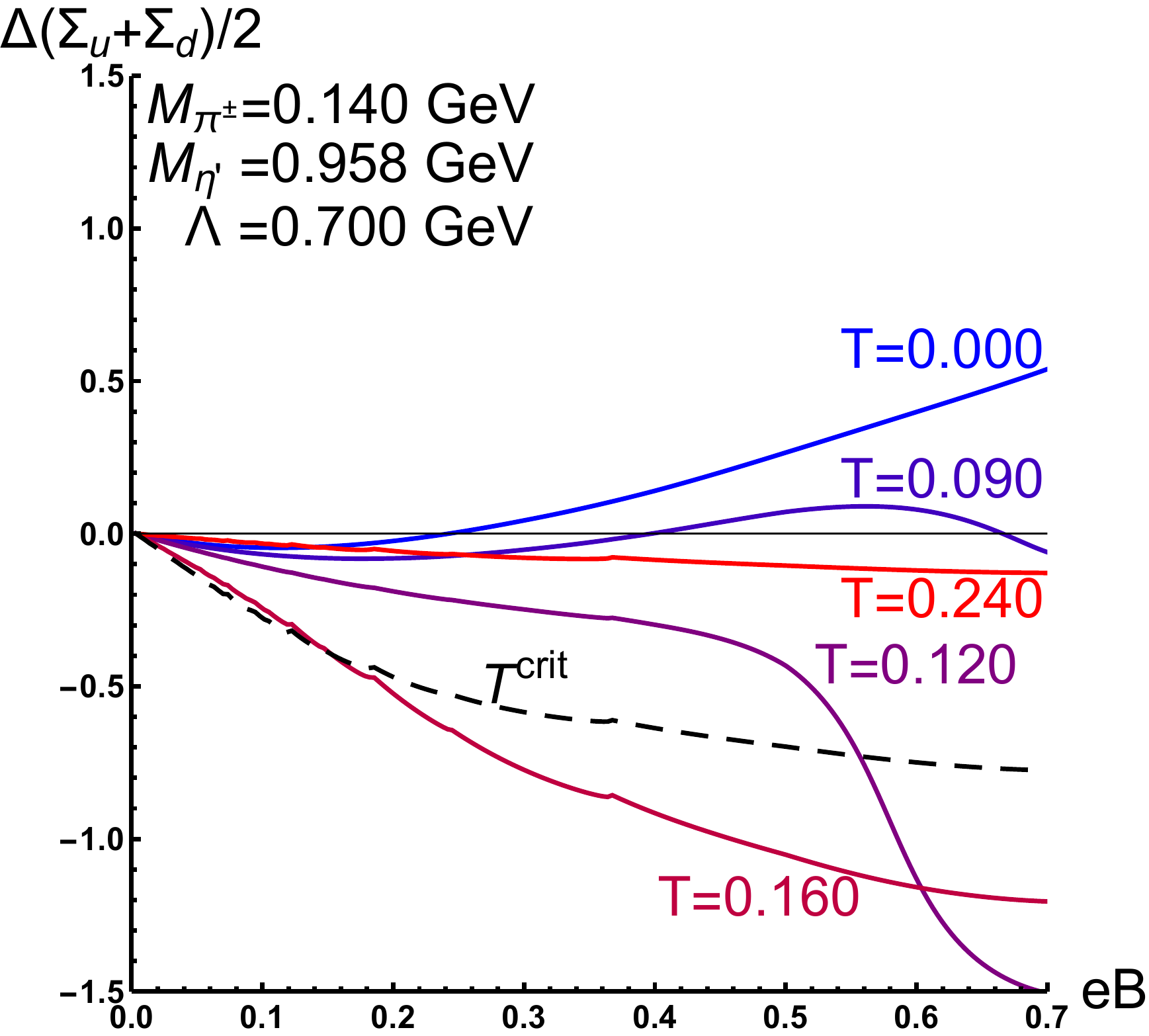}}
\subfigure[]{\label{grafSigmaIMCMdMslQCDMPi0140MEtal0958MEta0548PC}\includegraphics[width=0.32\textwidth]{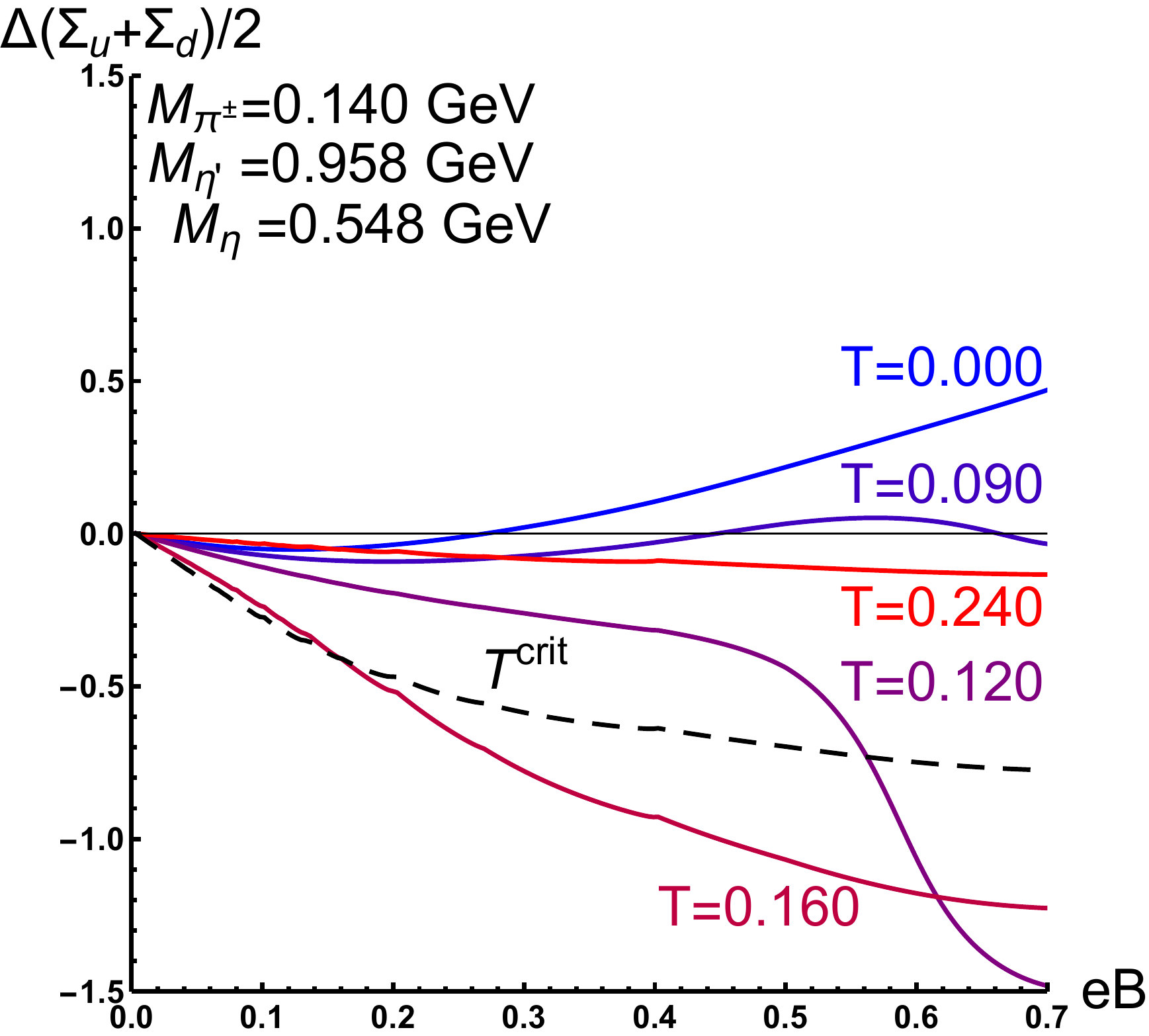}}
\caption{Renormalized chiral condensate change with magnetic field at different temperatures as given by our model calculations for sets from Table \ref{ParameterSetsI}.}
\label{CondvsBIMC}
\end{figure*}

This variation of the magnetic field dependence of the renormalized chiral condensate change at different temperatures is depicted in Fig. \ref{CondvsBIMClQCD} (LQCD data taken from \cite{Bali:2012zg}) and \ref{CondvsBIMC} (the results obtained with the parameter sets from Table \ref{ParameterSetsI}). The quantity displayed in Figs. \ref{FaixasCondensadoIMClQCD} and \ref{CondvsBIMC} is given by:
\begin{align}
&\Sigma_{i}\left(B, T\right) = \nonumber\\
&\frac{2m_i}{ M_\pi^2 F_\pi^2}
\left(
\langle \overline{\psi}_i\left(B,T\right)\psi_i\left(B,T\right)\rangle-
\langle \overline{\psi}_i\left(0,0\right)\psi_i\left(0,0\right)\rangle
\right)+1,\nonumber\\
&\Delta \Sigma_i\left(B,T\right)=\Sigma_i\left(B,T\right)-\Sigma_i\left(0,T\right).
\end{align}
Here $F_\pi$ is the pion decay constant ($F_\pi=86~\mathrm{MeV}$ for the LQCD data and $F_\pi=77.088$, $87.7928$, $96.8062$ and $99.3428~\mathrm{MeV}$ for sets $a$, $b$, $c$ and $d$ from Table \ref{ParameterSetsI}), $M_\pi$ the pion mass ($M_\pi=135~\mathrm{GeV}$ for the LQCD data and $M_\pi=140~\mathrm{GeV}$ for our sets) and $i$ is the flavor of the quark.

As can be seen in Figs. \ref{CondvsBIMC} the behavior is well reproduced qualitatively by our parameter sets. It should be noted however that while on the LQCD data the nonmonotonic behavior (the \emph{bump}) is more marked just below the pseudocritical transition temperature for partial chiral restoration in the light sector ($T^{\chi_l}_c=0.158~\mathrm{GeV}$ for the LQCD data), in our case this behavior occurs for lower temperatures. The pseudocritical temperatures for our sets (determined as the inflexion points of the average light quark condensates) are $T^{\chi_l}_c=0.170$, $0.175$, $0.180$ and $0.182~\mathrm{GeV}$ for sets $a$, $b$, $c$ and $d$ from Table \ref{ParameterSetsI} respectively. The marked \emph{bump} occurs in ours sets for a temperature around half the pseudocritical temperature or even slightly lower. There is also a small decrease for lower magnetic fields which is not observed in LQCD data points and which is more pronounced at larger cutoff.


\begin{figure*}
\center
\subfigure[]{\label{grafTclvsBlQCD}\includegraphics[width=0.32\textwidth]
{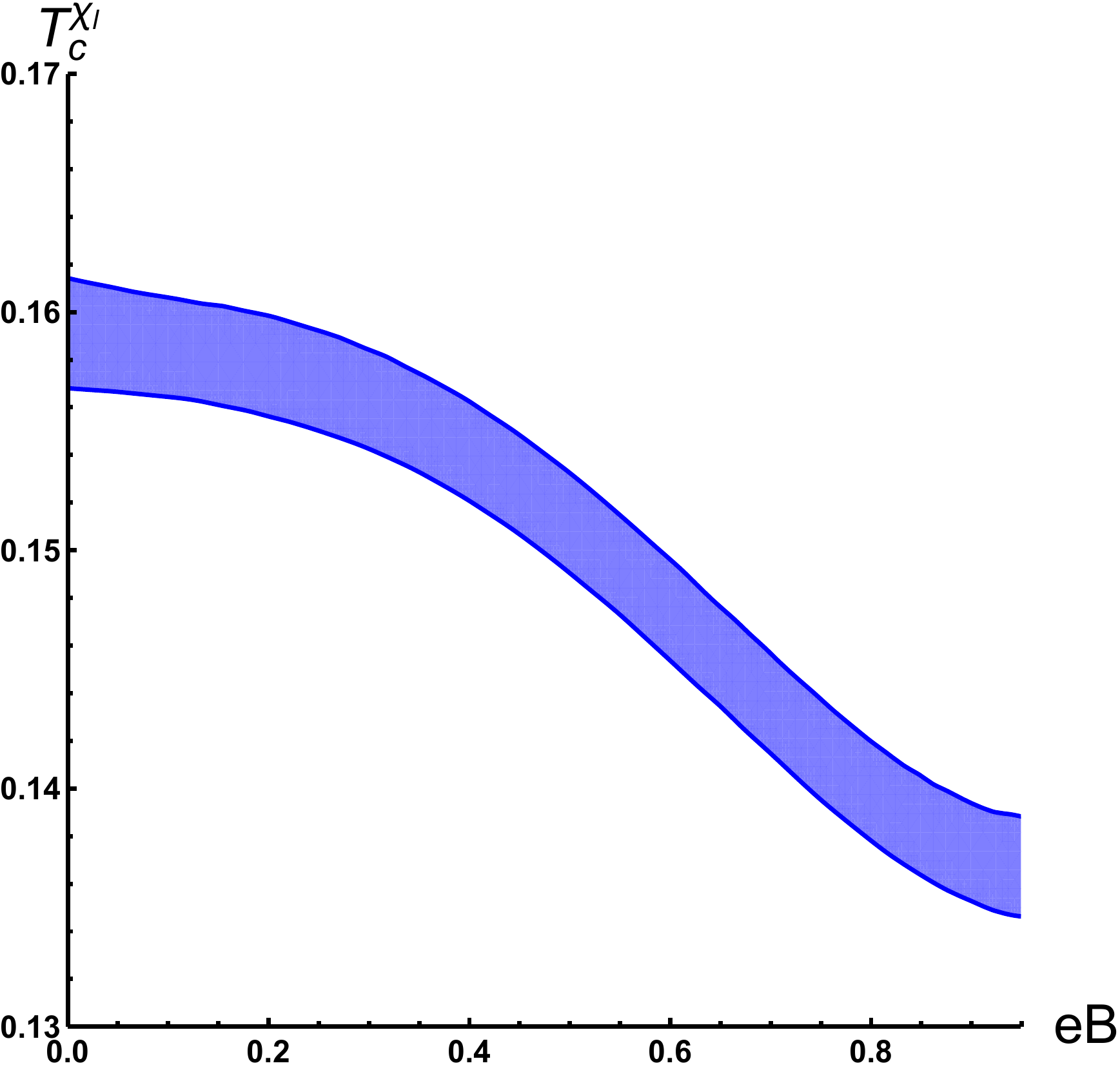}}
\subfigure[]{\label{grafTcChilvsBMPi0140MEtal0958Lambda050006000700MEta0548}\includegraphics[width=0.32\textwidth]
{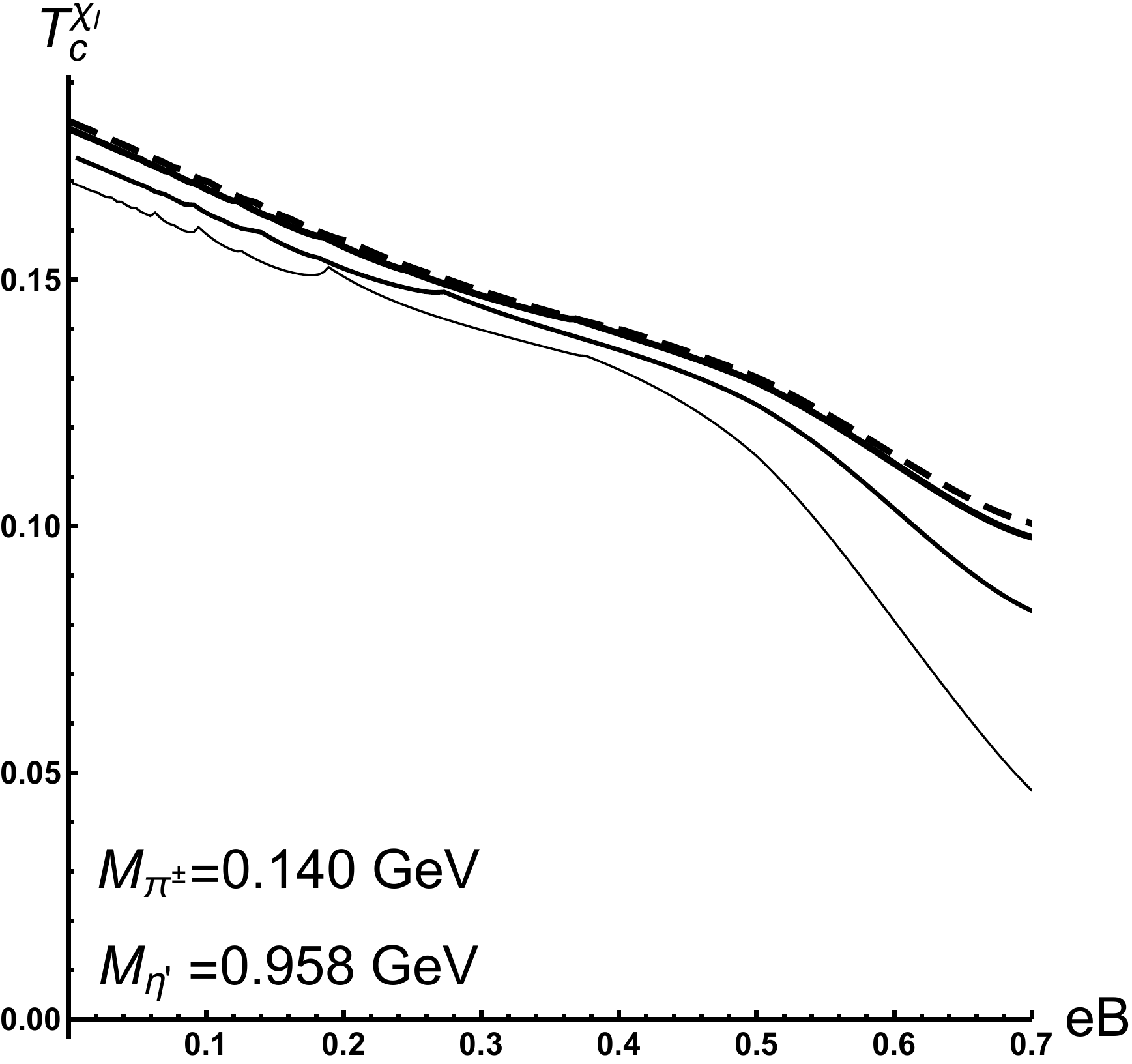}}
\caption{Pseudocritical temperature for the partial restoration of chiral symmetry in the light sector. On the left-hand side the LQCD results reported in \cite{Bali:2011qj} whereas on the right-hand side the results obtained with our model are presented.  On the latter, the full lines in increasing order of thickness correspond to: $\Lambda=0.500~\mathrm{GeV}$, $0.600~\mathrm{GeV}$ and $0.700~\mathrm{GeV}$. The thick dashed line corresponds to set $d$ in Table \ref{ParameterSetsI} which reproduces $M_\eta=0.548$. All sets reproduce $M_{\pi}=0.140~\mathrm{GeV}$ and $M_{\eta'}=0.958~\mathrm{GeV}$. }
\label{grafTcIMC}
\end{figure*}

The decrease in the pseudocritical temperature for partial chiral defined by the inflection point in the average light quark condensate 
\begin{align}
\left.\frac{\mathrm{d}^2  }{\mathrm{d} T^2} \frac{1}{2}\left(\langle\overline{\psi}_u \psi_u\rangle+\langle\overline{\psi}_d \psi_d\rangle\right)\right|_{T=T^{\chi_l}_c}=0
\end{align}
is depicted in Fig. \ref{grafTcChilvsBMPi0140MEtal0958Lambda050006000700MEta0548} where one can see that the temperature is larger and its behavior becomes more linear for larger cutoffs. The decrease of the critical temperature with increasing magnetic field strength is however much more pronounced in our model calculations. As one can see in Fig. \ref{grafTcChilvsBMPi0140MEtal0958Lambda050006000700MEta0548} by $e B=0.7~\mathrm{GeV^2}$ the critical temperature has dropped to a value which is approximately half of its value for vanishing field while, for the same magnetic field strength, LQCD show a reduction of only $\sim 10\%$ approximately. It should also be noted that while for larger cutoff an approximate linear response to magnetic field is obtained in our model, for the LQCD simulation the slope changes with magnetic field, hinting at an inflexion point.  

One should stress that both the magnetic field dependence of the pseudocritical temperature for partial chiral restoration as well as the temperature dependence of the magnetic field change induced renormalized chiral condensate are easily modified by considering further extensions of the model such as the inclusion of a Polyakov potential which mimics gluon dynamics and would modify the finite temperature aspects of the model.
Also the idea that extensions of the model to include higher order interaction terms, for instance as in \cite{Osipov:2005tq, Osipov:2006ns,Osipov:2012kk, Osipov:2013fka,Hiller:2008nu}, or go beyond mean field corrections as in \cite{CamaraPereira:2020ipu,CamaraPereira:2020xla,PhysRevC.81.065205,PhysRevD.91.065017}, could possibly fix these issues and deserves to be explored.

\section{Conclusions}\label{sec4}
\label{Conclusions}

In this work we have studied the implications of the magnetic field dependence of the dynamical quark masses on the \emph{up}, \emph{down} and \emph{strange} sectors (as reported in \cite{Endrodi:2019whh}) by using the 't Hooft extended Nambu--Jona-Lasinio model. 
Our study can be broken down into two separate steps:
\begin{itemize}
\item at vanishing magnetic field we developed four parameter sets, all of which reproduce the pion and the eta prime physical meson masses as well as the vacuum dynamical masses of the quarks; for three of them we imposed a choice of cutoff whereas for the remaining set we chose to reproduce the eta meson mass.
\item at vanishing magnetic field we used the variation in the $down$ and $strange$ quarks dynamical masses to fit the coupling strengths of the model interactions [$G(B)$ and $\kappa(B)$ for the NJL four-quark interaction and the 't Hooft flavor determinant six-quark interaction respectively] while keeping the other parameters frozen at their vacuum value. Then, we studied the inverse magnetic catalysis phenomenology having arrived at an acceptable qualitative agreement with LQCD data both for the decrease in the pseudocritical temperature for the partial chiral symmetry restoration in the $up$ and $down$ sector with increasing magnetic field as well as the different magnetic field dependent behaviors at different temperatures.
\end{itemize}
We found for the four-fermion coupling, $G(B)$, an overall similar behavior to the ones reported in previous works. 
Regarding the six-fermion 't Hooft coupling, $\kappa$, our results showed that the absolute value of $\kappa$ increases as $B$ increases.
To our knowledge this is the first work with a six-fermion 't Hooft flavor determinant running with the magnetic field. 
We also verified that the obtained parameter sets, as well as their magnetic field dependencies, show promising results which can be easily applied to further studies, for instance, the Polyakov loop dynamics (see \cite{Fukushima:2003fw, Megias:2003ui, Megias:2004hj, Roessner:2006xn, Ghosh:2007wy,Fu:2007xc,Costa:2008dp, Fukushima:2008wg, Bhattacharyya:2010wp, Moreira:2010bx, Stiele:2016cfs, Bhattacharyya:2016jsn}). This same procedure can also be applied to any new LQCD data as well as for different regularization procedures.

\section*{Acknowledgments}
This work was supported by a research grant under Project No. PTDC/FIS-NUC/29912/2017, funded by national funds through FCT (Fundação para a Ciência e a Tecnologia, I.P, Portugal) and cofinanced by the European Regional Development Fund (ERDF) through the Portuguese Operational Program for Competitiveness and Internationalization, COMPETE 2020, by national funds from FCT, within the Projects No. UID\slash 04564\slash 2019 and No. UID\slash 04564\slash 2020. This study  was  financed  in  part  by Coordena\c c\~{a}o  de Aperfeiçoamento de Pessoal  de  N\'{\i}vel Superior-(CAPES-Brazil)-Finance  Code 001. We would like to thank G. Endr\"odi for his availability in supplying the data points from the LQCD study reported in \cite{Endrodi:2019whh}. T.E.R. thanks the support and hospitality of CFisUC and 
acknowledges Conselho Nacional de Desenvolvimento Cient\'{\i}fico e Tecnol\'{o}gico (CNPq-Brazil) and  Coordena\c c\~{a}o  de
Aperfei\c coamento  de  Pessoal  de  N\'{\i}vel  Superior   (CAPES-Brazil) for   PhD grants at different periods of time.

\bibliographystyle{apsrev4-1}
\bibliography{GKfit_lQCDMi}
\end{document}